\documentclass[fleqn,11pt]{article}

\usepackage{arxiv}
\usepackage{cite}
\usepackage{filecontents}
\usepackage{titletoc}

\usepackage[utf8]{inputenc}
\usepackage[T1]{fontenc}

\usepackage{gensymb}

\usepackage{hyperref}

\usepackage{color}

\usepackage{amsmath,amsfonts,amssymb}
\usepackage{graphicx}
\usepackage{booktabs}

\usepackage[left=3cm,right=3cm,top=3.25cm,bottom=3.25cm,headheight=12pt,letterpaper]{geometry}

\title{Event-Chain Monte-Carlo Simulations of Dense Soft Matter Systems} 

\author{
  Tobias A. Kampmann \\
  Physics Department \\
  TU Dortmund University \\
  \texttt{tobias.kampmann@udo.edu} \\
   \And
  Jan Kierfeld \\
  Physics Department \\
  TU Dortmund University \\
  \texttt{jan.kierfeld@udo.edu} 
  \And
  David M\"uller \\
  Physics Department \\
  TU Dortmund University 
  \And
  Clemens Franz Vorsmann \\
  Physics Department \\
  TU Dortmund University 
  \And
  Lukas Paul Weise \\
  Physics Department \\
  TU Dortmund University 
}

\newcommand{\comment}[1]{}

\usepackage{xcolor}

\usepackage{url,hyperref,lineno,microtype,subcaption}
\usepackage{textcomp}
\usepackage[onehalfspacing]{setspace}
\usepackage{bm}

\renewcommand{\vec}[1]{\mathbf{#1}}
\usepackage{microtype}
\usepackage{textcase}

\newcommand{\liftr}{p^{\text{lift}}(\textcolor{rred}{\text{red}}\to\textcolor{ggreen}{\text{green}})}
\newcommand{\liftg}{p^{\text{lift}}(\textcolor{ggreen}{\text{green}} \to \textcolor{rred}{\text{red}})}

\newcommand{\Ered}{d\mathcal{E}_{\textcolor{rred}{\text{red}}}}
\newcommand{\Egreen}{d\mathcal{E}_{\textcolor{ggreen}{\text{green}}}}

\newcommand{\lamrg}{ \lambda_{\textcolor{rred}{\text{red}} \to \textcolor{ggreen}{\text{green}}} }
\newcommand{\Eblue}{d\mathcal{E}_{\textcolor{bblue}{\text{blue}}}}

\newcommand{\lamg}{ \lambda_{\textcolor{ggreen}{\text{green}}} }

\definecolor{rred}{HTML}{c04040}
\definecolor{ggreen}{HTML}{54c040}
\definecolor{bblue}{HTML}{4164c0}

\begin{document}

\flushbottom
\maketitle

\begin{abstract}

We discuss the rejection-free event-chain  Monte-Carlo 
algorithm and several applications to dense soft matter
systems. Event-chain Monte-Carlo is an alternative to
standard local Markov-chain Monte-Carlo schemes, which
are based on  detailed balance, for example the well-known
Metropolis-Hastings algorithm. Event-chain Monte-Carlo
is a Markov chain Monte-Carlo scheme that uses so-called
lifting moves to achieve global balance without
rejections (maximal global balance).
It has been originally developed for hard
sphere systems but is applicable to many soft matter
systems and particularly suited for dense soft matter
systems with hard core interactions, where it gives
significant performance gains
compared to a local Monte-Carlo simulation. The algorithm can be
generalized to deal with soft interactions and with
three-particle interactions, as they naturally arise,
for example, in bead-spring models of polymers with bending rigidity.
We present results for polymer
melts, where the event-chain algorithm can be used for an efficient
initialization. We then move on to large systems of semiflexible
polymers that form bundles by attractive interactions and can serve as model
systems for actin filaments in the cytoskeleton. The event chain
algorithm shows that these systems form networks of bundles which coarsen
similar to a foam. Finally, we present results on liquid crystal systems, where
the event-chain algorithm can equilibrate large systems containing additional
colloidal disks very efficiently, which reveals the
parallel chaining of disks. 

\end{abstract}

\newpage

\section{Event-Chain Monte-Carlo Algorithm}

Since its first application to a hard disk system \cite{metropolis1953},
Monte-Carlo (MC) simulations have been applied to virtually all types of 
models in statistical physics, both on-lattice and off-lattice. 
MC samples microstates $\{a,  b, \ldots\}$ of a thermodynamic ensemble 
statistically according to their Boltzmann weight
$\pi_a = \exp \left(- E_a / k_BT \right)$
(in the following, we use thermal units, $k_BT \equiv 1$).
One advantage of MC schemes over, for example molecular dynamics
simulations, is that 
it only requires knowledge of 
microstate  energies $E_a$ rather than interaction forces. 
In its simplest form, the Metropolis-Hastings
algorithm  \cite{metropolis1953,Hastings1970},
a MC simulation is easily implemented for any system by 
offering moves from a microstate $a$ to a  microstate $b$ of
the system and accepting or rejecting these moves  
according to the Metropolis rule for the Boltzmann distribution, i.e.,
based on the energy difference between states $\Delta E_{a\rightarrow b}
= E_ b-E_a$.
The standard Metropolis rule is defined by the acceptance probability 

\begin{align}
  p^{\text{Metr}}(a\rightarrow b)
    &=\min\left(1,\frac{\pi_b}{\pi_a}\right)
               =\min(1,\exp(-\Delta E_{a\rightarrow b}))
      = \exp(-[\Delta E_{a\rightarrow b}]^+)
      \label{eq:metropolis} \\
 \text{with } [x]^+ &\equiv \max(0,x) \nonumber 
\end{align}

for a move $a\rightarrow b$.
Together with the trial probability $p^\text{trial}(a\rightarrow b)$
that a move $a\rightarrow b$ is proposed we obtain the transition
rate as the product
$p(a\rightarrow b)=p^\text{trial}(a\rightarrow b)
p^\text{Metr}(a\rightarrow b)$.
The transition rates $p(a\rightarrow b)$ give rise to
probability currents $ J_{a\rightarrow  b}  =  \pi_a p(a\rightarrow b)$
from state $a$ to $b$.
Transition rates $p(a\rightarrow b)$ are given  per MC time;
likewise probability currents $J_{a\rightarrow  b}$
have units probability per MC time, which must not be identified
  with an  actual physical time.
The Metropolis rule is designed to fulfill
\emph{detailed balance} of probability currents between any
states $a$ and $b$, i.e.,

\begin{align}
  J_{a\rightarrow  b}  =  \pi_a p(a\rightarrow b)
  &=   \pi_b p(b\rightarrow a) = J_{b\rightarrow  a}.
  \label{eq:db}
\end{align}

The Metropolis rule   (\ref{eq:metropolis}) satisfies
detailed balance 
for the Boltzmann distribution
$\pi_a = \exp \left(- E_a \right)$
if moves are also offered with a
symmetric trial probability $p^\text{trial}(a\rightarrow b)
=p^\text{trial}(b \rightarrow a)$ (which is typically fulfilled
trivially for standard local MC moves).
A characteristic of the 
Metropolis rule (\ref{eq:metropolis})  is also that moves that are
energetically downhill, 
$\Delta E_{a\rightarrow b}<0$, are \emph{always} accepted
($p^\text{Metr}(a\rightarrow b)=1$). We point out that
detailed balance (for symmetric
trial probabilities) and this maximal acceptance rate for energetically
downhill moves determine the Metropolis rule for the Boltzmann
distribution uniquely. 
The Metropolis
acceptance rate for moves that are energetically uphill is
exponentially small, however, resulting in frequent rejections.

Typical local MC moves, such as single spin flips in spin systems 
or single particle moves in off-lattice systems of interacting particles, 
are often  motivated by the actual dynamics of the system.
Sampling  with local moves 
can become slow, however, in certain
physically relevant situations, most notably, 
close to a critical point, where large correlated regions 
exist, or in dense systems, where acceptable moves become rare.

Cluster algorithms deviate from the local Metropolis MC rule and
construct MC moves of large non-local clusters, ideally, in
a way that the MC move of the cluster is performed rejection-free. For lattice
spin systems, the Swendsen-Wang \cite{Swendsen1987} and Wolff \cite{Wolff1989}
algorithms are the most important cluster algorithms with enormous performance
gains close to criticality, where they reduce the dynamical exponent governing
the critical slowing down in comparison to local MC simulations.
These cluster algorithms still fulfill detailed balance but based on a 
non-trivial trial probability that derives from the cluster construction
 rules.

The event-chain (EC) MC algorithm introduced by Krauth {\it et al.}
\cite{bernard2009} has been developed to decrease the autocorrelation time
in 
hard disk or sphere systems. It performs rejection-free displacements of
several spheres in a single  MC move
(see Fig.\ \ref{fig:hard_spheres}) and can, therefore, 
also be  classified as a cluster algorithm. The basic idea is to perform a
displacement $\ell$ in a \emph{billiard}-type  fashion,
transferring the displacement to
the next disk upon a collision, which is called an \emph{event}
and gives the algorithm its name 
 (see Fig.\ \ref{fig:hard_spheres}).
As opposed to cluster algorithms such as
the Swendsen-Wang or Wolff algorithms,
ECMC algorithms do no longer fulfill detailed balance but only
 global balance.

 In the following, we will give an introduction into
 the ECMC algorithm starting with the example of simple hard sphere
 systems. We then discuss how the algorithm realizes the
 general concept of lifting moves, which is employed
 in order to achieve  global balance. Finally, we
 review the adaptation and
 generalization of the ECMC algorithm from athermal systems with
 hard interactions to systems with
 \emph{soft} potentials specified by an interaction energy.
Many soft matter systems
can be constructed from hard spheres as basic building blocks, eventually
with additional interactions, for example spring-like connections
to form a polymer chain. 
We  discuss in some detail
several applications of the EC algorithm in dense soft matter systems,
namely polymer systems, liquid crystal systems
as well as mixed systems such as liquid crystal colloids.

\subsection{EC Algorithm For Hard Spheres}

\begin{figure}
\centering
 \includegraphics[width=0.8\linewidth]{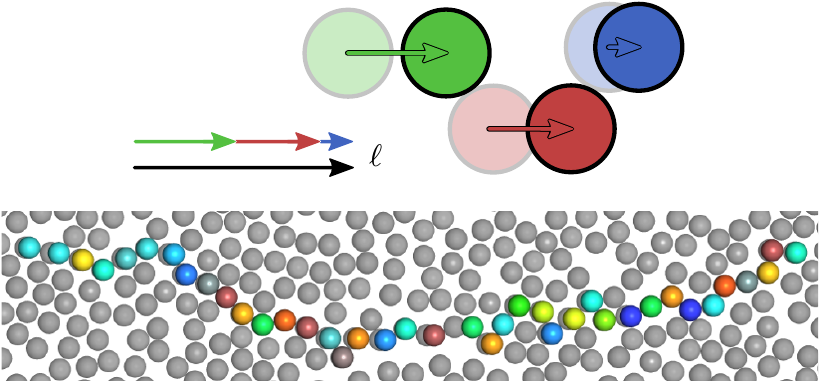}
 \caption{The upper
   scheme illustrates the \emph{billiard}-type
   construction of an EC with a total displacement
   length $\ell$ and with three participating spheres. 
   The lower picture shows the example of a typical   EC in a
   dense hard-sphere system: the coherent movement of long 
   line-like clusters in each ECMC move
   accelerates the sampling by nearly two orders of
   magnitude.  }
 \label{fig:hard_spheres}
\end{figure}

We consider the conceptually
simple system of hard  (impenetrable)
disks (of diameter $\sigma$)
in two dimensions in order to  introduce the ECMC algorithm.
 Hard disks are the epitome of a system that is easily described and
 quickly implemented in a (naive) simulation \cite{metropolis1953},
 but exhibits a non-trivial phase transition that has been
 debated for a long time \cite{Strandburg1988}.
 The 
 two-dimensional melting
 phase transition is believed to be a two-step Kosterlitz-Thouless 
 melting via an intermediate hexatic phase
 \cite{kosterlitz1973,Halperin1978,Young1979}, which is, however,
 difficult to confirm  unambiguously  in simulations
 \cite{Mak2006,krauth2011}.

In a hard disk system,  an EC move is constructed  by
first  selecting a random  starting particle, an EC  displacement direction,
and a 
 total displacement length
 $\ell$. Both direction and  total displacement length $\ell$  are
 adjustable parameters of the algorithm
  (see Fig.\ \ref{fig:hard_spheres}); the  sampling of EC directions 
 of directions (uniform or non-uniform)
 will determine whether detailed or  global
 balance is fulfilled, while
 the displacement length $\ell$ can be adjusted for performance.
 We start the EC by  moving the chosen 
 particle in the  EC direction, which is only possible
 until it touches another
 particle after a certain displacement $r$, which will be typically 
 much smaller than the intended displacement $\ell$ in a dense system.
 Then, instead of declining such a move as in local MC schemes,
 the  displacement $r$ of the first particle is subtracted from the
 initial total displacement length and the remaining displacement $\ell-r$
 is
 carried over to the hit particle, which we attempt to displace next
 by the remaining distance $\ell-r$.
 If also the displacement direction is carried over to the hit particle
 the EC is called \emph{straight}; this is the case we will
 focus on in the following and which is illustrated in Fig.\
 \ref{fig:hard_spheres}.
 This procedure of lifting the EC to the next particle upon
 collision 
 continues until no displacement length is left. Then an EC has been
 constructed that moved as a line-like cluster without
 the possibility of rejection, and we start over by choosing
 a new starting  particle and a new random starting direction
 to construct the next EC with the same  total displacement length $\ell$.

 The total displacement length
$\ell$ is an adjustable algorithm parameter, which determines
the average number of disks $n_{\text{EC}}$ that are moved in an EC move
(see Fig.\ \ref{fig:hard_spheres}).
This number is given by $n_{\text{EC}} \sim \ell/\lambda_0$,
where $\lambda_0$ denotes the mean free path; 
in general,  the efficiency of the  EC algorithm  depends crucially
on $n_{\text{EC}}$.
For smaller $\ell$, the efficiency decreases and 
approaches traditional local MC (corresponding to the case
$\ell<\lambda_0$).
For very large $\ell$, ECs comprise large parts of the system
and give rise to motion similar to a collective translation of disks,
which is also inefficient.

We note that the EC algorithm will be very well suited to simulate
  dense systems but does not allow to  simulate maximally dense, i.e.,
  jammed systems since an EC can not displace any particle if
  all particles touch each other. In jammed hard disk systems,
  the mean free path $\lambda_0$ approaches zero
  such that the number of participating disks
  $n_{\text{EC}}$ diverges, and the EC does no longer terminate or
    essentially comprises the entire system resulting in collective
    translation. This will happen
    for $n_{\text{EC}} \sim \ell/\lambda_0 > N$.
    In a dense hard disk system of area $L^2$ and close to the
    close-packing volume fraction $\eta_{hcp}= \pi/2\sqrt{3}$, we have
    $\lambda_0 \approx \sigma (\eta_{hcp} - \eta)/2\eta_{hcp}$  
    resulting in a more concrete condition 
    $\ell \lesssim (\eta_{hcp}-\eta) L^2/\sigma$ for the
    EC algorithm to remain effective close to close-packing.

We will show in the next section that 
the EC algorithm ensures \emph{global balance}, regardless of how
the  EC  displacement direction is chosen.
If the EC displacement directions are chosen randomly the straight
EC algorithm 
also satisfies \emph{detailed balance} over many EC events. 
The most performant version of the straight EC algorithm, however,
the so-called
xy-version or irreversible straight EC algorithm, breaks detailed balance
and uses only displacements along
coordinate axes and only in positive direction.
Breaking detailed balance but preserving global balance leads
to performance gains. Ergodicity must
also be ensured, which is done
via  periodic boundary conditions for the irreversible EC algorithm;
moreover, several computations can be simplified
for the xy  EC algorithm. In Ref.\ \citenum{engel2013}, the
irreversible EC simulations are roughly $70$ times faster than a
simulation employing local MC moves.
We can obtain similar performance gains in dense soft matter systems.

\subsection{Balance Conditions And Lifting For Hard Spheres}

The standard  Metropolis algorithm (\ref{eq:metropolis})
employs detailed balance to ensure stationarity of
the probability distribution of states by pairwise
balancing of  probability fluxes
between microstates in the  condition (\ref{eq:db}).
Since circular probability fluxes also fulfill the
stationarity requirement for the probability distribution, detailed balance is
not a  necessary condition and, often, performance can be gained
by  finding an algorithm that fulfills the weaker
condition of \emph{global balance},

\begin{align}
  \sum_{  b } J_{ b\rightarrow  a}
  &= \sum_b \pi_b p(b\rightarrow a) = \pi_a   \sum_c p({a\rightarrow c})
    =  \sum_c J_{a\rightarrow c}  
          = {\pi_a}
   \label{eq:GB}                                  
\end{align}

(where the sums are over all microstates),
which is simply the continuity equation for a  stationary
probability $\pi_a$, i.e.,  the Boltzmann distribution in our case.
While the last equality in (\ref{eq:GB}) is a trivial consequence of
normalization, the second equality is the actual global balance condition. 
All quantities in eq.\ (\ref{eq:GB}) are transition probabilities
\emph{per MC step}; we call the algorithmic time
measured in MC steps $\tau$ in the following. The algorithmic
time $\tau$ measured in MC steps  must not be identified with an actual
\emph{physical time} $t$ as for any MC simulation.

Global balance algorithms should become particularly performant if 
\emph{maximal global balance} is achieved, which
means that probability backflow is forbidden leading to
the additional constraint
\begin{align}
  J_{a\rightarrow b} \neq 0 \Rightarrow J_{ b\rightarrow a} = 0.
  \label{eq:mgb}
\end{align}
In particular, this implies a rejection-free  algorithm, i.e.,
$J_{a\rightarrow a} = 0$.  
Then we can also take  a 
\emph{hydrodynamic} point of view and see probability density as liquid mass
density, unidirectional transition probabilities as liquid velocities, and
the probability currents as liquid current densities.
Then, maximal global balance corresponds to  stationarity of the
  mass distribution in the presence of fluid flow, i.e., a divergence-free
  liquid current density.

\begin{figure}
\centering
 \includegraphics[width=0.8\linewidth]{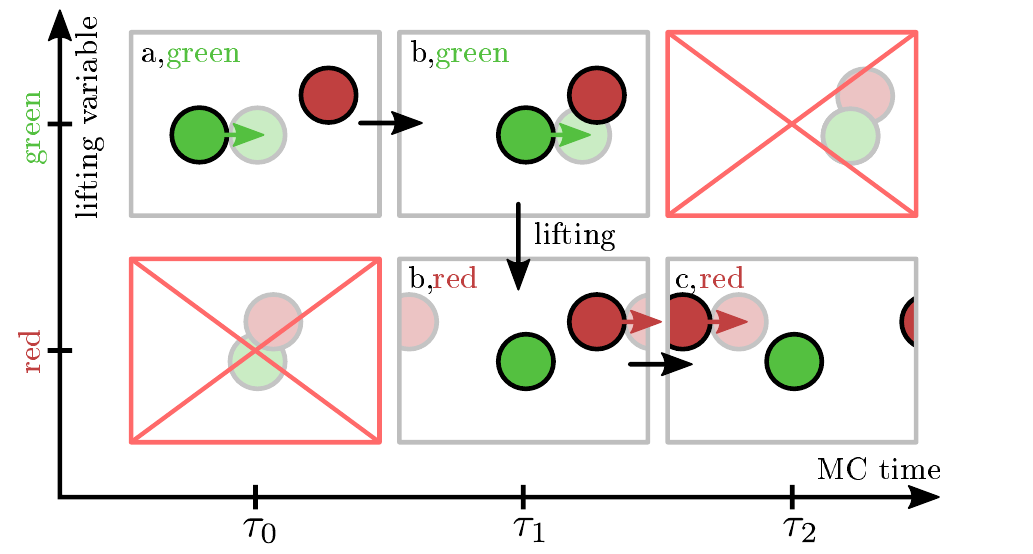}
 \caption{
   Scheme of the lifting formalism for hard spheres.
   The physical states are labeled with
   letters (a,b,c) and the lifting variable is indicated by
   color, i.e., the current
   active particle is indicated by a colored arrow.
   The EC  displacement direction is to the right. 
   Proposed infinitesimal displacements
are indicated by opaque spheres. In case of a physical
rejection at MC step $\tau_1$ the pivot particle is changed
and rejection is avoided.  Periodic
boundaries ensure ergodicity.
One way to read this scheme, is that the lower row is the
displacement-reversed version (except trivial translations) of the upper row
(obtained be reversing all displacements $dw\to -dw$),
resulting in a central symmetry, especially
{($c$,\textcolor{rred}{red})} $=$ {($a$,\textcolor{ggreen}{green})}.
 }
 \label{fig:lifting}
\end{figure}

In Fig.\ \ref{fig:lifting}, we show how maximal global balance is achieved
in EC algorithms for a hard sphere system.
In order to investigate balance conditions, it is often convenient
to decompose the EC into infinitesimal moves $dw$, which add up
to the total EC displacement  $\ell$ in the end
\cite{michel2014,michel2015,harland2017}. 
To ensure maximal global balance 
\emph{lifting} transitions are introduced, which change an
additional lifting variable that
enlarges the state space of the  system \cite{Diaconis2000}. 
For hard sphere systems, the additional lifting variable is simply the
number $i$ of the active/moving particle.
This will be the case throughout this article. 
The microstates are then characterized not only by the
positions of all particles but also by specifying the active
particle.

For infinitesimal EC moves, only two spheres can collide at once, and
it is sufficient to consider a 
simplified setting of only two hard spheres (\textcolor{ggreen}{green}
and \textcolor{rred}{red}) as shown in the cartoon in
Fig.\ \ref{fig:lifting}, where proposed infinitesimal displacements
are indicated by opaque spheres.
The physical states are  labeled with
   letters (a,b,c), and the lifting variable is indicated by
   color (\textcolor{rred}{red} or \textcolor{ggreen}{green}
   particle is active).
  \emph{Physical} transitions change the physical states by changing
   particle positions; this happens by moving the active sphere
in the EC  displacement direction
(i.e., to the right from {($a$,\textcolor{ggreen}{green})} at $\tau_0$ to
{($b$,\textcolor{ggreen}{green})} at $\tau_1$ in
Fig.\ \ref{fig:lifting}). If  a physical transition leads to an
overlap, instead of rejecting the move as in local MC,
we  perform a \emph{lifting} transition and
change the lifting variable with a lifting probability that
exactly equals the rejection probability of the local MC rule in the EC
algorithm.
In Fig.\ \ref{fig:lifting}), this means that  the active particle is changed
to the  \textcolor{rred}{red} sphere if a collision is proposed at
$\tau_1$.
Then the \textcolor{rred}{red} active particle continues to move
(to the right from {($b$,\textcolor{rred}{red})} to
{($c$,\textcolor{rred}{red})}.

In this way, rejections are avoided,
and the entire rejection
probability flow is \emph{redirected} to a lifting move probability
flow, 
which ensures maximal global balance if
all physical configurations are equally probable, i.e.,
if the Boltzmann distribution for hard spheres holds:
\begin{itemize}
\item
  The total physical
  inflow  $J^{\text{phys}}_{\rm a,\textcolor{ggreen}{green}
    \rightarrow b,\textcolor{ggreen}{green}}$
   to {($b$,\textcolor{ggreen}{green})}  equals
    the  rejected physical flow
    $J^{\text{phys}}_{\rm b,\textcolor{ggreen}{green}\rightarrow coll}$
    (that would lead
  to collisions on the upper right picture)
  because it involves moving the same sphere by the same
  distance $dw$ and because physical
  states $a$ and $b$ are equally probable.
\item 
  By construction of the EC algorithm, all collisions give rise to lifting
   such that 
   the lifting flow $J^{\text{lift}}_{\rm b,\textcolor{ggreen}{green}
     \rightarrow b,\textcolor{rred}{red}}$
  equals the rejected physical flow
  $J^{\text{phys}}_{\rm b,\textcolor{ggreen}{green}\rightarrow coll}$.
  This leads to 
  $J^{\text{phys}}_{\rm a,\textcolor{ggreen}{green}\rightarrow
    b,\textcolor{ggreen}{green}} =
  J^{\text{lift}}_{\rm b,\textcolor{ggreen}{green}\rightarrow b,
    \textcolor{rred}{red}}$, so far. 
\item
  Apart from  a translation,  the states $a$ and $c$ and the
  \emph{forbidden} overlapping
  configurations (upper right and lower left pictures in
  Fig.\ \ref{fig:lifting})
  are identical (please note the periodic boundary
  conditions).
\item
  Therefore, the  missing physical inflow
  $J^{\text{phys}}_{\rm coll \rightarrow b,\textcolor{rred}{red}}$
  (from the forbidden
  overlapping configuration  on the lower left picture)  into
 state {($b$,\textcolor{rred}{red})} equals the  rejected physical flow
 $J^{\text{phys}}_{\rm b,\textcolor{ggreen}{green}\rightarrow coll}$ and, thus,
 the lifting flow $J^{\text{lift}}_{\rm b,\textcolor{ggreen}{green}
   \rightarrow b,\textcolor{rred}{red}}$.
\item
 Again,  the missing  physical
   inflow   $J^{\text{phys}}_{\rm coll \rightarrow b,\textcolor{rred}{red}}$  into
 state {($b$,\textcolor{rred}{red})}
    equals
    the   physical outflow
    $J^{\text{phys}}_{\rm b,\textcolor{rred}{red}\rightarrow c,\textcolor{rred}{red}}$
    from {($b$,\textcolor{rred}{red})}
  because it involves moving the same sphere by the same
  distance $dw$ and because physical states $b$ and $c$ are equally probable.
\end{itemize}
All in all, we obtain 
\begin{align*}
  J^{\text{phys}}_{\rm a,\textcolor{ggreen}{green}
  \rightarrow b,\textcolor{ggreen}{green}}
  &= J^{\text{lift}}_{\rm b,\textcolor{ggreen}{green}
    \rightarrow b,\textcolor{rred}{red}}
    = J^{\text{phys}}_{\rm b,\textcolor{rred}{red}
    \rightarrow c,\textcolor{rred}{red}},
 \end{align*}
 which are  exactly the maximal global balance conditions
 for states {($b$,\textcolor{ggreen}{green})} and
{($b$,\textcolor{rred}{red})} 
proving maximal global balance for any EC move
on  a state space that is extended by the lifting
 variable denoting the active particle.
 Global balance  holds, regardless of how
the  EC  displacement direction or displacement length is chosen.
If the reverse EC displacement directions are offered
with the same  probability (for example by drawing EC directions
completely uniformly in all directions)
the resulting  EC algorithm 
also satisfies  detailed balance over the course of many
EC moves.

For pair interactions between hard spheres the lifting variable is the
particle number. We can also introduce hard walls to the system, which 
represent additional  steric one-particle interactions.
For such one-particle interactions, the lifting variable will not be
the particle number (as there are no interaction partners, the active particle
must stay the same) but the
EC direction itself.  It can be shown that
classical reflection of the EC direction upon collision with
the  hard wall will ensure maximal
global balance. Analogously to the forbidden overlapping
configurations Fig.\ \ref{fig:lifting}, both the rejected outflow
in a  collision with a wall and the missing inflow from forbidden
configurations are  exactly equal to the lifting flow from
reflected EC directions.

\subsection{Soft  Interaction Energies}

In many applications other than pure hard core systems, we have to deal with
systems containing hard core repulsion alongside with other \emph{soft}
interaction energies $\mathcal{E} = \mathcal{E}(\vec{r}_1,...,\vec{r}_N)$
that can depend
on all particle positions $\vec{r}_n$ in a
canonical ensemble. These are not handled by
the hard sphere EC scheme described  so far. 
One naive strategy to handle an additional interaction is the following:
to ensure ergodicity, an EC is limited to a total displacement $\ell$,
i.e., the sum of all particle displacements, after which a random new particle
is chosen as active and a new isotropic direction is set. Interpreting a whole
EC as a single  MC move,
an  additional interaction will cause an energy
change $\Delta \mathcal{E}$ when a hard sphere EC (which
ignores the additional soft interaction) is executed.
To properly sample the
Boltzmann distribution in the presence of the additional interaction, the 
EC as a whole is then accepted or rejected by a standard Metropolis
filter \eqref{eq:metropolis}.

This strategy of handling additional interaction energies by re-introduction
of Metropolis sampling and, thus, rejections will surely
decrease the efficiency of the simulation. All advantages of the
EC algorithm are effectively lost. 
There is, however, a way to  include arbitrary
$\mathcal{N}$-particle
interaction energies  into  the EC framework.
Within the lifting formalism it
is possible to extend the algorithm from hard spheres to arbitrary pair
potentials and continuous spin models by employing  factorized Metropolis
filters \cite{peters2012,michel2014,nishikawa2015}; this concept
can be further generalized to $\mathcal{N}$-particle interactions
with $\mathcal{N}>2$ according to Ref.\ \citenum{harland2017}. 
In principle, these 
expansions of the EC algorithm presented in the following
can be reduced to our  simple example of a two-sphere collision
from Fig.\ \ref{fig:lifting} by 
generalizing  the notion of a \emph{collision} \cite{peters2012}.
Effectively, we assign a \emph{virtual} hard sphere radius to
each interaction, which determines which configurations are energetically
forbidden and trigger a generalized collision.
This can be done by introducing the
so-called \emph{rejection  distance},
which is further explained in Fig.\  \ref{fig:2-EC}.

 In a pure hard sphere simulation, a sphere is
continuously displaced until it hits another sphere which triggers a lifting
event (or the total displacement reaches $\ell$).
We can resolve all  events into unique binary
collisions or rejections, which then leads
to a simple lifting probability flux
during each event (see Fig.\ \ref{fig:lifting}).
When a particle is displaced in the presence of
continuous interaction energies, it is not clear which interaction
causes the
rejection, since the rejection in the standard Metropolis filter is based on
the sum of all changes in all interaction energies (pair or
many-particle interactions). In fact, the
rejection is caused by all interactions \emph{at once}. By using an
infinitesimal factorized Metropolis filter one can handle each interaction
independently.

Each particle $i$ has a set $S_i$ of interactions with other particles;
each interaction energy
$\mathcal{E} \in  S_i$ depends on the position $\vec{r}_i$ of particle $i$ and
on the  positions of other particles in the set $M_{\mathcal{E}}$
of interaction partners  participating in $\mathcal{E}$.
Pairwise interactions, such as elastic springs
$\mathcal{E}=E_{\rm spring}(ij)$ in
a bead-spring polymer or a van-der-Waals interaction
between particles, are interactions with one other particle $j$
($M_{E_{\rm spring}(ij)}=\{j\}$; all particles $j\neq i$ can be
interaction partners);
three-particle interactions such as a bending energy in a polymer are
interactions $\mathcal{E}=E_{\rm bend}(jik)$
with two other particles $j,k$ that form two neighboring bonds with $i$
and thus define a bond angle ($M_{E_{\rm bend}(ijk)}=\{j,k\}$;
all pairs $jk$ with $j\neq i, k \neq i$ can be
interaction partners).

In the following, we consider again infinitesimal displacements
$dw$ of particle $i$ in the EC direction. 
Such an infinitesimal move changes the configuration from $b$ to $a$
and
leads to the total energy change $dE_{i,b\rightarrow a}$,
which is the sum of all changes in  interactions in $S_i$,
 $dE_{i,b\rightarrow a} = \sum_{\mathcal{E} \in S_i}
    d\mathcal{E}_{i,b\rightarrow a}$.

We construct a global balance algorithm starting from a
 Metropolis filter for the 
acceptance of physical moves based on the energy $dE_{i,b\rightarrow a}$.
   For soft interactions  we  switch, however,  from the 
 standard Metropolis filter \eqref{eq:metropolis} to  a factorized one
 \cite{peters2012,michel2014,nishikawa2015}, where
 the Boltzmann weights for the sum $dE_{i,b\rightarrow a} =
 \sum_{\mathcal{E} \in S_i}
 d\mathcal{E}_{i,b\rightarrow a}$ are factorized and the Metropolis
 acceptance rule is applied to each factor separately.
This is equivalent to 
 switching the sum and maximum operation on the energy changes
regarding each interaction,

\begin{align}
  p^\text{fact}(i,b\rightarrow a)
  &    =\exp \left({-\sum_{\mathcal{E} \in S_i}
        [d\mathcal{E}_{i,b\rightarrow a}]^+}\right)
    \approx 1-\sum_{\mathcal{E} \in S_i} [d\mathcal{E}_{i,b\rightarrow a}]^+.
\label{eq:FactMet}
\end{align}

It will be important that
the factorized Metropolis filter still has the  detailed balance property
(\ref{eq:db}).
The 
probability of rejecting the next infinitesimal step $1- p^\text{fact}$ is
proportional to the sum of each infinitesimal energy change of each
interaction. If one interaction causes a rejection, the whole move is
rejected. Since we employ infinitesimal steps, only one interaction can reject
at once. In analogy to the hard sphere case, these rejection events can be
seen as a generalized collision events. 
In other words, the factorized Metropolis filter uniquely assigns
a rejection to one specific interaction at the expense of an increased
rejection rate, because energy increase from one interaction cannot be
compensated by an energy decrease of another interaction.

The factorized Metropolis filter can now be used to construct a maximal
global
balance algorithm on an enlarged state space
by redirecting all physical rejection events into lifting
events. This is done  analogously to the hard sphere
case outlined above.
If a rejection should take place according to
the factorized Metropolis filter, we perform a
compensating lifting move instead.
We lift to one of the particles that participate in the interaction
that caused the rejection.
For a pair interaction this already determines
the particle to lift to uniquely, for many-body interactions an additional
rule is needed to select the particle to lift to properly \cite{harland2017}.
Therefore, we first discuss the simpler case of pair interactions
and show how we can realize maximal global balance.

\subsubsection{Pair Interactions}

\begin{figure}
\centering
 \includegraphics[width=0.85\linewidth]{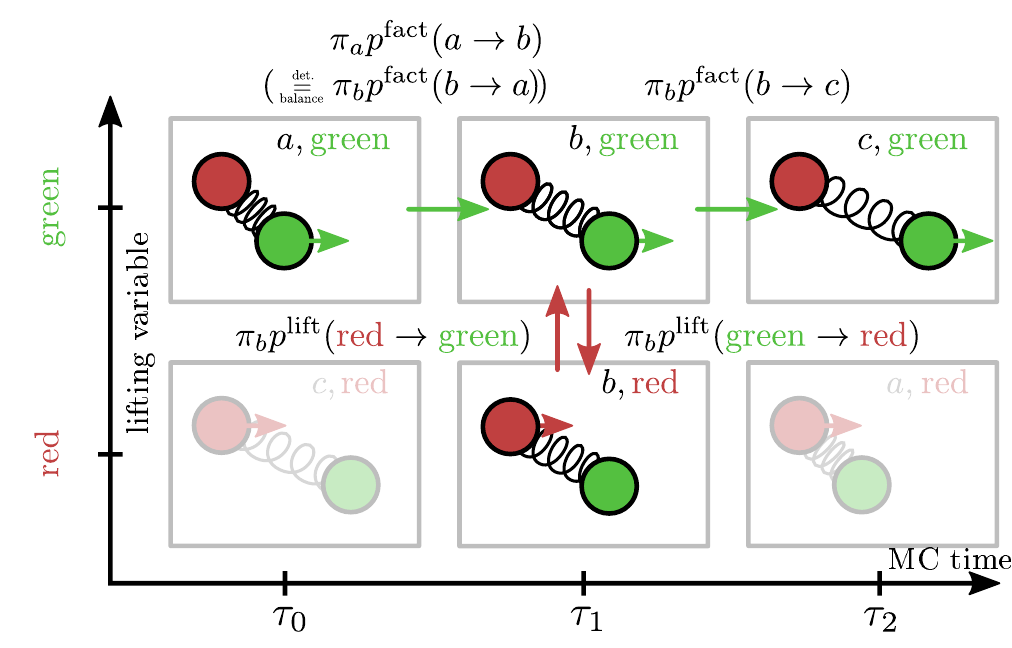}
 \caption{
  Scheme of the lifting formalism for a continuous pair
  interaction (notation see Fig.\ \ref{fig:lifting}).
  Using eq.\ \eqref{eq:FactMet} in the global balance
   condition (\ref{eq:GB}) for incoming and outgoing
     currents for state $b$ and dividing by
   $\pi_b$ yields $ \liftr -[d\mathcal E_{b\to c}]^+ = \liftg
   -[d\mathcal E_{b\to a}]$, where $d\mathcal E_{b\to c} = -d\mathcal
   E_{b\to a} \equiv d\mathcal E$ because of translational invariance.
   $[d\mathcal E]^+=0$ leads to $\liftr =0$, yielding
   $\liftg =[-d\mathcal E]^+$ and vice versa, which assures
    maximal global balance.
    Detailed balance for physical transitions (green)
    ensures time reversibility, so
   that a reversed EC would redo the original one, while an
   EC itself satisfies maximal global balance. For clarity we completed
     the lower row to show  translational symmetry.
   }
 \label{fig:spring_EC}
\end{figure}

For states in the enlarged state  space, we use  the notation
$a_i$  for a physical
state $a$ with an active particle $i$. 
In the following we try to move the active particle $i$ by an infinitesimal
step $dw$ along the EC direction $\vec{d}$
resulting in a physical move from $a_i$ into $b_i$. 
For pairwise interactions
$\mathcal{E}(ij)$ (i.e., $S_i = \{\mathcal{E}(ij)| j\neq i\}$)
the transition rate is given by the factorized Metropolis filter 

\begin{align}
  p^{\text{phys}}(a_i\rightarrow b_i)
  &= p^{\text{fact}}(i,a\rightarrow b)
    =\exp \left({-\sum_{\mathcal{E} \in S_i}
       [d\mathcal{E}_{a_i\rightarrow b_i}(ij)]^+}\right)
    \approx 1-\sum_{\mathcal{E} \in S_i}
    [d\mathcal{E}_{a_i\rightarrow b_i}(ij)]^+ \, .
    \label{eq:FactMetpair}
\end{align}

This transition rate is decreased
by rejections if a move has $d\mathcal{E}_{a_i\rightarrow b_i}(ij)>0$, i.e.,
is energetically uphill.
If  $p^{\text{phys}}(a_i\rightarrow b_i)>0$ because
configuration $b_i$ can be reached by a move $dw$ of particle $i$ from $a_i$,
it follows that $p^{\text{phys}}(b_i\rightarrow a_i)=0$ because the
EC direction is fixed, and we cannot reach $a_i$ from $b_i$ by moving
$i$ into the same direction $dw$ (we would need to move into the
opposite direction $-dw$). Therefore, physical transitions
  are unidirectional.

Because only one pair interaction  $\mathcal{E}(ij)$
   can reject at once in the factorized
   Metropolis filter, it is sufficient to consider the two interacting
   particles $i$ and $j$ in deriving the lifting probabilities
   $p^\text{lift}_{\mathcal{E}(ij)}(b_i\rightarrow b_j|a_i\rightarrow b_i)$
    from  global balance,
    see Fig.\ \ref{fig:spring_EC}.
    We want to apply the   global balance condition  (\ref{eq:GB}) to
    incoming   physical and lifting flows to state $b_i$.
    Using detailed balance of the
factorized Metropolis filter we obtain for the physical inflow from $a_i$ to
$b_i$
    
    \begin{align}
        J^{\text{phys}}_{ a_i\rightarrow  b_i}
  &= \pi_{a_i}  p^{\text{fact}}(i,a\rightarrow b)
    = \pi_{b_i}  p^{\text{fact}}(i,b\rightarrow a)
  = {\pi_{b_i}}
    \left(1-
        [-d\mathcal{E}_{a_i\rightarrow b_i}(ij)]^+\right),
    \label{eq:Jphysbi}
\end{align}

 Lifting moves   from $b_j$ to $b_i$ are triggered
  by a move $dw$ of particle $j$ into configuration $b_j$
from a different configuration $c_j$, which is obtained from $b_j$
by displacing particle $j$ by  $-dw$
($c_j$ is analogous to the forbidden configuration on the lower left
in Fig.\ \ref{fig:lifting}).

\begin{align}
  J^\text{lift}_{b_j\rightarrow b_i}
     &=  \pi_{b_j}
    p^\text{lift}_{\mathcal{E}(ij)}(b_j\rightarrow b_i|c_j\rightarrow b_j)
    =       
       \pi_{b_i}
       p^\text{lift}_{\mathcal{E}(ij)}(b_j\rightarrow b_i|c_j\rightarrow b_j)
   \label{eq:Jliftbi}                                
\end{align}

 with $\pi_{b_i} = \pi_{b_j} = \pi_b/N$ by symmetry
in a $N$-particle system. 
Using the global balance condition  (\ref{eq:GB}) for
flows to state $b_i$,

\begin{align}
  J^{\text{phys}}_{ a_i\rightarrow  b_i} +
  J^\text{lift}_{b_j\rightarrow b_i}  &= {\pi_{b_i}},
    \label{eq:GB2}
\end{align}

and dividing by $\pi_{b_i}$ we obtain
the lifting probability as

\begin{align}
   p^\text{lift}_{\mathcal{E}(ij)}(b_j\rightarrow b_i|c_j\rightarrow
  b_j) 
  &=
    [-d\mathcal{E}_{a_i\rightarrow b_i}(ij)]^+
    = [d\mathcal{E}_{c_j\rightarrow b_j}(ij)]^+.
\label{eq:liftpairback}
\end{align}

  The last equality holds for a translationally invariant
pair interaction, where 
$d\mathcal{E}_{c_j\rightarrow b_j}(ij)=
-d\mathcal{E}_{a_i\rightarrow b_i}(ij)$.
In the other lifting direction, this implies 
\begin{align}
  p^\text{lift}_{\mathcal{E}(ij)}(b_i\rightarrow b_j|a_i\rightarrow
  b_i) &=
         [d\mathcal{E}_{a_i\rightarrow b_i}(ij)]^+,
         \label{eq:liftpair}
\end{align}
which  leads to
  a rejection-free algorithm because the rejection
  probability of the physical moves  (\ref{eq:FactMetpair})
  is exactly redirected into a lifting probability.
We also conclude that $p^\text{lift}_{\mathcal{E}(ij)}(b_j\rightarrow b_i) >0$
 requires $p^\text{lift}_{\mathcal{E}(ij)}(b_i\rightarrow b_j) =0$, i.e.,
 also lifting transitions are unidirectional proving maximal global
 balance.

\begin{figure}[t]
\centering
 \includegraphics[width=0.85\linewidth]{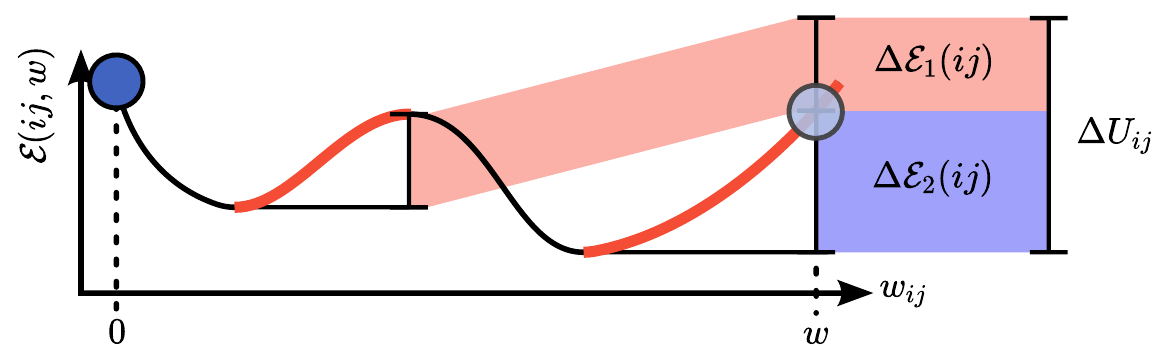}
 \caption{ Event-driven approach to determine the rejection distance for a
    pair
   interaction between $i$ and $j$. Condition \eqref{eq:faster_than} has to be
   solved for the rejection distance $w_{ij}$ at which an event/lifting
   occurs. The black curve is the interaction energy along the EC
   direction.  Only on red segments with increasing energy a rejection can
   occur.  Among all pair interactions,
    the rejection will be assigned to the interaction with
   the smallest rejection distance.
 }
 \label{fig:2-EC}
\end{figure}

 Global balance is established
 for each move $a_i\rightarrow  b_i$ independently
 This generalizes the picture from
Fig.\ \ref{fig:lifting} to soft interactions.
In fact, for hard spheres $[d\mathcal{E}_{a_i\rightarrow b_i}(ij)]^+=0$ before
a collision and $[d\mathcal{E}_{a_i\rightarrow b_i}(ij)]^+=\infty$ at collision
such that lifting to the colliding particle $j$ happens with probability
one upon collision, exactly as in Fig.\ \ref{fig:lifting}.

For an efficient implementation, an event-based approach is chosen
\cite{peters2012,michel2014}, in which we determine the
distance to the next rejection. This rejection distance
corresponds to the  distance to the next collision for hard spheres. 

The probability $p(w_{ij})$
to reject and lift from  particle $i$ to particle $j$
because of the interaction $\mathcal{E}(ij)$ and 
after moving 
a distance in the interval $[w_{ij},w_{ij}+dw]$ is given by the probability
to not encounter
a rejection/lifting up to $w_{ij}$ and then lift with
probability $p^\text{lift}_{\mathcal{E}(ij)}(w_{ij})dw
=  [d\mathcal{E}_{a_i\rightarrow b_i}(ij,w_{ij})]^+$,
see eq.\ (\ref{eq:liftpair}).
As a result, we obtain a Poisson-type distribution for the
rejection distance $w_{ij}$ of particle $i$ caused by
interactions 
with particle $j$,  

\begin{align}
  p(w_{ij}) &= \exp\left(-\int_0^{w_{ij}}dw  \left[
              \partial_w \mathcal{E}(ij,w)\right]^+  \right)
              \left[\partial_w \mathcal{E}(ij,w_{ij})\right]^+,
      \label{eq:pw}
\end{align}  

which depends on all \emph{uphill} energy differences
caused by particle $j$  if particle $i$ is moved, see Fig.\ \ref{fig:2-EC}.
Interpreting  $-\partial_w \mathcal{E}(ij,w)$ as force onto particle $i$
exerted by particle $j$, the rejection length distribution depends on
the line  integral over all \emph{opposing} forces.

Transforming the probability distribution (\ref{eq:pw}),
we find that the  distribution of the \emph{usable} (i.e., uphill)
energy
$\Delta U_{ij} =
\int_0^{w_{ij}} dw \left[ \partial_w \mathcal{E}(ij,w)\right]^+ $ 
 is a simple exponential. 
It follows that the  lifting  distance $w_{ij}$
caused by particle $j$ can be determined with the proper
distribution by drawing a
probability $u_{ij}= p(\Delta U_{ij})$  from a uniform
distribution in $[0,1]$, which translates into an 
exponentially
distributed $\Delta U_{ij}=-\ln u_{ij}$,
and by solving the equation 

\begin{align}
  - \ln u_{ij} = \Delta U_{ij}
  = \int_0^{w_{ij}} dw \left[ \partial_w \mathcal{E}(ij,w)\right]^+ 
\label{eq:faster_than}
\end{align}

for $w_{ij}$. This is illustrated in  Fig.\ \ref{fig:2-EC}.
Because all interaction energies $\mathcal{E}(ij)$  with different partners
$j$
are independent, rejection probabilities from the factorized
  Metropolis filter are additive by construction, and we have a Poissonian
 rejection length distribution,
we can determine a rejection distance for
 all possible interaction energies, and 
 the shortest distance triggers the next  lifting event.
 This is analogous to the first-reaction method in
 a Gillespie algorithm \cite{Gillespie2007}.
 In this sense, the rejection distance is a virtual collision
 radius, which can be assigned to each interaction, and the
 shortest collision radius triggers a generalized  collision. 
This procedure based on (\ref{eq:faster_than})
gives a simple and fast event-driven ECMC algorithm for arbitrary
pair potentials.

\subsubsection{$\mathcal{N}$-Particle Interactions}

In many applications, also  three-particle
interactions occur. This happens, in particular, for extended objects, such as
rods or polymers, which can be described by bead-spring models. One prominent
example are semiflexible polymer chains with a bending energy.  Because the
local bending angle involves three neighboring beads in a discrete model, the
bending energy is a three-particle intra-polymer interaction in terms of bead
positions.

In Ref.\ \citenum{harland2017}, the  EC algorithm has been generalized
to three- and many-particle
interactions, thus broadening the range of applicability of rejection-free EC
algorithms considerably. For $\mathcal{N}$-particle interactions
$\mathcal{E}$, there are
$\mathcal{N}-1$ interacting particles to which the EC can lift to avoid
rejections.

In order to redirect the rejection probability
$[d\mathcal{E}_{a_i\rightarrow b_i}]^+$ from physical moves
into lifting moves and, thus, a rejection-free algorithm,
lifting  to the set $M_{\mathcal{E}}$
 of $\mathcal{N}-1$
interaction partners  should be done with the
analogous probability

\begin{align}
  p^\text{lift}_{\mathcal{E}}(b_i\rightarrow b_{M_{\mathcal{E}}}|a_i
  \rightarrow b_i)
      =  [d\mathcal{E}_{a_i\rightarrow b_i}]^+,
\label{eq:liftN}
\end{align}

as for pairs, see eq.\ (\ref{eq:liftpair}).
We need, in addition, a set of
conditional lifting probabilities $\lambda_{ij}$ to assure
maximal global balance.  The $\lambda_{ij}$ specify the probabilities to lift
to  one of the 
$\mathcal{N}-1$ interaction partners $j\in M_{\mathcal{E}}$
($\sum_{j\in M_{\mathcal{E}}} \lambda_{ij}=1$);
the total probability to lift to $j$ becomes

\begin{align}
p^\text{lift}_{\mathcal{E}}(b_i\rightarrow b_j|a_i\rightarrow b_i)
  &=  [d\mathcal{E}_{a_i\rightarrow b_i}]^+\lambda_{ij}.
    \label{eq:pliftNlambda}
\end{align}

\begin{figure}[t]
\centering
 \includegraphics[width=0.95\linewidth]{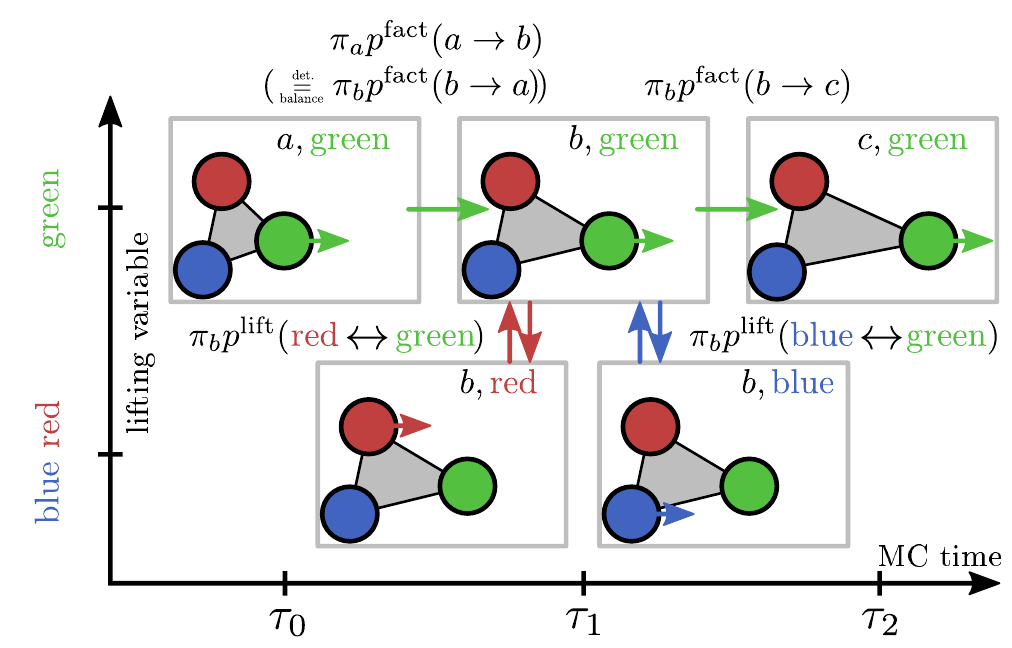}
 \caption{
   Conditional lifting in a system with three-particle interaction.
   The conditional lifting probabilities can be
   derived from the balance condition. Analogously to
   Fig.\ \ref{fig:lifting} (i.e., using eq.\ \eqref{eq:FactMet}
   and dividing by $\pi_b$) and decomposing the
   lifting probabilities according to eq.\ \eqref{eq:pliftNlambda}
   and choosing $\lamrg \equiv \lamg$,
   we get $ [-\Egreen]^+  = [\Ered]^+ \lamg +  [\Eblue]^+ \lamg $
   from the inflowing probability fluxes, leading to $\lamg = [-\Egreen]^+ /
   ([\Ered]^+ + [\Eblue]^+)$. Note that this can be
   extended to match eq.\ \eqref{eq:lambdaij}.
 }
 \label{fig:3_part}
\end{figure}

We first consider the important case of   
a three-particle interaction  (between particles $i$, $j$, $k$), see also
Fig.\ \ref{fig:3_part}.
Analogously to eq.\ (\ref{eq:liftpairback}),
the global balance condition  (\ref{eq:GB}) applied to 
flows to state $b_i$
requires a backward lifting probability

\begin{align}
   p^\text{lift}_{\mathcal{E}(ijk)}(b_j\rightarrow b_i|c_j\rightarrow
  b_j)+
  p^\text{lift}_{\mathcal{E}(ijk)}(b_k\rightarrow b_i|c_k\rightarrow
  b_k)
  &
  =  [-d\mathcal{E}_{a_i\rightarrow b_i}(ijk)]^+.
\label{eq:liftpairback3}
\end{align}

Inserting (\ref{eq:pliftNlambda}) this results in the condition

\begin{align}
  [d\mathcal{E}_{c_j\rightarrow  b_j}(ijk)]^+\lambda_{ji} +
  [d\mathcal{E}_{c_k\rightarrow  b_k}(ijk)]^+\lambda_{ki} =
  [-d\mathcal{E}_{a_i\rightarrow b_i}(ijk)]^+.
  \label{eq:GB3}
\end{align}

As for pairs, lifting from particle $i$ to $j$
will only take place ($\lambda_{ij}>0$)
if $d\mathcal{E}_{a_i\rightarrow b_i}>0$ giving
rise to a rejection of $a_i\rightarrow b_i$, and if 
the corresponding $d\mathcal{E}_{c_j\rightarrow b_j}<0$ because of eq.\
(\ref{eq:GB3}).
Translational invariance of the interaction implies
$d\mathcal{E}_{a_i\rightarrow b_i} + d\mathcal{E}_{c_k\rightarrow  b_k}
+d\mathcal{E}_{c_j\rightarrow  b_j}=0$. 
Writing out eq.\ (\ref{eq:GB3}) analogously
also for lifting to $j$ and $k$, we can 
determine all conditional lifting probabilities
uniquely \cite{harland2017},

\begin{align}
  \lambda_{ij}    &= \Theta(d\mathcal{E}_{a_i\rightarrow b_i})
            \frac{\left[-d\mathcal{E}_{c_j\rightarrow b_j}\right]^+}
  {[d\mathcal{E}_{a_i\rightarrow b_i}]^++[d\mathcal{E}_{c_k\rightarrow
                    b_k}]^+} ,
  \label{eq:lambdaij3}
 \end{align}

 where $\Theta(x)$ is the Heaviside step function.

 For arbitrary $\mathcal{N}$, the global balance condition (\ref{eq:GB3})
   generalizes to

   \begin{equation}
  \sum_{j\in  M_{\mathcal{E}}}
  [d\mathcal{E}_{c_j\rightarrow b_j}]^+\lambda_{ji} =
    [-d\mathcal{E}_{a_i\rightarrow b_i}]^+ \, ,
  \label{eq:GBN}
\end{equation}

For  $\mathcal{N}>3$, the choice of conditional lifting probabilities
is not unique  \cite{harland2017}.
They can be fixed by the additional
  requirement $\lambda_{ji} \equiv \lambda_i$, that they
  are independent of the
  particle $j$ we are lifting from  as  in (\ref{eq:lambdaij3}).
This leads to the simple general result

\begin{align}
  \lambda_{ij} &= \Theta(d\mathcal{E}_{a_i\rightarrow b_i})
                \frac{\left[-d\mathcal{E}_{c_j\rightarrow b_j}\right]^+}
                 {[d\mathcal{E}_{a_i\rightarrow b_i}]^++
                 \sum_{k\in  M_{\mathcal{E}}\backslash  \{i,j\}}
                [d\mathcal{E}_{c_k\rightarrow b_k}]^+ }.
  \label{eq:lambdaij}
 \end{align}

Regardless of the number of interacting
particles $\mathcal{N}$ in an interaction in $S_i$,
the first step of calculating the rejection distance does not
differ and is given by (\ref{eq:faster_than}). 
  For $\mathcal{N}=2$, the lifting destination when a
    rejection occurs is unambiguous, hence $\lambda_{ij} =1$.
  For $\mathcal{N} \geq 3$, the probability to
lift from the active particle $i$ to one of the other interacting particles
$j\in  M_{\mathcal{E}}$
is also
proportional to $\lambda_{ij} \propto \left[-d\mathcal{E}_j\right]^+$
(see eq.\ (\ref{eq:lambdaij})),
which can be
interpreted as the \emph{force} which is exerted on the particle $j$
by the
interaction. Note that the conditional lifting probability vanishes when the
particle, one would lift to, will increase the energy when moving along the
current EC direction. We want to highlight that due to symmetry the
sum of all forces must vanish $ \sum_{j}dE_j = 0$ and $i$ can only trigger
an event when $dE_i > 0$ holds, which implies that for at least one other
particle  $dE_j < 0$ must hold. In short, one most probably lifts to the
particle, which decreases the interaction energy the fastest.

\subsubsection{One-Particle Interactions}

Finally, soft  external one-particle  potentials $\mathcal{E}(i)$
frequently occur in soft matter systems, 
for example, as 
confining potentials, gravity,
or external electric fields for charged particles.
Such one-particle
potentials are handled analogously to hard walls. Because there
are no interaction partners the lifting  variable is the
EC direction $\vec{d}$. The lifting probability

\begin{align}
  p^\text{lift}_{\mathcal{E}}(\vec{d}\rightarrow \vec{d}'|a_i
  \rightarrow b_i)
      =  [d\mathcal{E}_{a_i\rightarrow b_i}]^+,
\label{eq:lift1}
\end{align}

for lifting the EC direction to a reflected direction $\vec{d}'$
during an attempted move $a_i\rightarrow b_i$. The reflection
can be performed
with respect to the equipotential surface of $\mathcal{E}(i)$, i.e.,
by lifting to  $\vec{d}' = (\bm{1}-2\bm{P})\vec{d}$, where 
$\bm{P}$ is the projection operator onto the
$\vec{\nabla}_i\mathcal{E}(i)$-direction.
This has been shown to satisfy global balance
\cite{peters2012}.

\subsection{ECMC Algorithm}

In summary, an ECMC simulation, which we use for soft matter systems,
can be structured as follows. Starting from a random particle $i$ in a random
EC direction $\vec{d}$ and with a certain EC total displacement $\ell$,
the rejection distance $w_{ij}$
for each interaction $\mathcal{E}\in S_i$
is calculated from  eq.\ \eqref{eq:faster_than} and
as shown in Fig.\ \ref{fig:2-EC}. The shortest distance determines which
interaction caused the rejection. If the interaction consists of more than two
particles, the conditional lifting probabilities $\lambda_{ij}$
are calculated from  eq.\ (\ref{eq:lambdaij}) and
as  illustrated in Fig.\ \ref{fig:3_part}.
Then we lift to the corresponding interacting   particle $j$ after
the corresponding rejection distance $w_{ij}$.
For one-particle interactions causing the
rejection we lift the EC direction by reflection.

\begin{enumerate}
\item
   \textbf{choose} a random EC direction $\vec d$, a total  displacement $\ell$ and  an active particle $i$ ,
 \item  \textbf{calculate} the minimal rejection distance
   $r_ {M_{\mathcal{E}}} = \min_{\mathcal{E'}\in S_i} w_{i M_{\mathcal{E'}}}$
   for each $\mathcal{E}$
   according to eq.\ \eqref{eq:faster_than},
 \item 
   if rejection is triggered by a three- or more  particle
   interaction:\\
   \textbf{select} $j \in  M_{\mathcal{E}}$ by applying conditional
   lifting probabilities  from  eqs.\ (\ref{eq:lambdaij3}) or
   (\ref{eq:lambdaij})
   (see Fig.\ \ref{fig:3_part}),
\item \textbf{move} particle $i$: $\vec{r}_i = \vec{r}_i + r \vec d$, 
\item \textbf{lift} to rejecting particle $i \to j$ (or to
  new reflected EC direction $\vec{d}'$ if rejection is triggered
  by a one-particle interaction), 
\item \textbf{let} $\ell = \ell -  r$,
\item \textbf{if} $\ell=0$ goto 1 \textbf{else} goto 2 
\end{enumerate}

\subsection{Additional ECMC Simulation Features}

We want to highlight some useful properties of an ECMC
simulation. First, using the factorized Metropolis filter each interaction is
independent, making it easy to implement the algorithm in a highly modular
manner.

Furthermore, the rejection distance distribution and lifting statistics is
intimately linked to the pressure in the system. Without loss of generality,
let the EC direction be in $x$-direction, then the pressure can be
expressed by \cite{michel2014}

\begin{align}
  \frac{P}{\rho k_B T} =
  \left \langle \frac{x_{\text{final}}-x_{\text{initial}}}{\ell}
  \right\rangle \, ,
  \label{eq:pressure}
\end{align}

where $\langle \ldots \rangle$ denotes an ensemble average over multiple ECs
with length $\ell$. Rewriting
$x_{\text{final}}-x_{\text{initial}} = \ell + \sum_{\text{events}}(x_j-x_i)$,
where $x_{i,j}$ are the $x$-coordinates of particles $i$ and $j$,
one sees that the excess length per EC length determines interaction
specific contributions to the pressure of an ideal gas, i.e., attractive
interactions with a negative excess length decrease the pressure,
and repulsive interactions
increase the pressure. This procedure does not need any additional
computations. For hard spheres, the pressure is measured far more
efficient than with the usual detour via the contact theorem.

Every configuration visited is weighted properly even if a hard sphere
collision just happened. This can be exploited by special MC moves. This is
not unconditionally true in a molecular dynamics simulation
where, for instance, the statistics is skewed if 
measurements are taken on particle collisions.
Inclusion of special many-particle
MC moves can be beneficial for particular systems. 
For  polymer melt simulations, we introduce
a bead-swapping move \cite{kampmann2015}, which switches
positions of two hard spheres upon collision
obeying a standard metropolis filter for the spring
energies involved, as explained later.

Many simulation techniques based on single particle moves
can benefit from massive
parallelization if the simulation system can be efficiently
divided into independent pieces.
For ECMC algorithms, massive  parallelization will not be
optimal because of the cluster nature of EC moves \cite{kampmann2015b}.
In Ref.\ \citenum{kampmann2015b},
a parallelization scheme for the EC algorithm was introduced,
and extensive tests for correctness and efficiency were performed
for the hard sphere system in
two dimensions.
For optimal parameters the algorithm
achieves a speed up by a factor of roughly the number of cores used compared
to a sequential EC simulation. For parallelization we use a spatial
partitioning approach into simulation cells.  We analyzed the performance
gains for the parallel EC algorithm and find the criterion
$\ell \sim {L\lambda_0}/{\sqrt{n}\sigma} $ for an optimal degree of
parallelization, where $n$ is the number of parallel threads, $L$ the system
size, and $\eta$ the
occupied area fraction. If the number $n$ of parallel threads is chosen too
large (massive parallelization),
this choice for $\ell$ will drop below the optimal window
$\ell < 10 \lambda_0$.
Because massive parallelization is not optimal, 
the parallel EC algorithm will
be best suited for commonly available multicore CPUs with shared
memory. This ECMC parallelization scheme can  also be
applied to other soft matter systems such as
polymer melts \cite{kampmann2015}.

\section{Applications}

We now briefly present some soft matter systems where we successfully
employed ECMC algorithms, namely a dense polymer melt, a
network-forming system of polymers with a
short range attraction, hard needle models of liquid crystal and
liquid crystal colloids from a mixture of needles and hard spheres.
In all of these systems the ECMC algorithm speeds up simulations considerably 
at high particle densities as compared to standard local MC simulations.
The simulations for these soft matter applications have been
performed within the highly modular  {\tt polyeventchain} framework,
which is particularly suited for soft matter applications
and will be presented in detail elsewhere.
 {\tt Jellyfish} provides a
similar EC framework with a focus on  all atom simulations
\cite{Faulkner2018,Hoellmer2020}.

\subsection{Initialization And Simulation Of Polymer Melts}

\begin{figure}
\centering
 \includegraphics[width=0.95\linewidth]{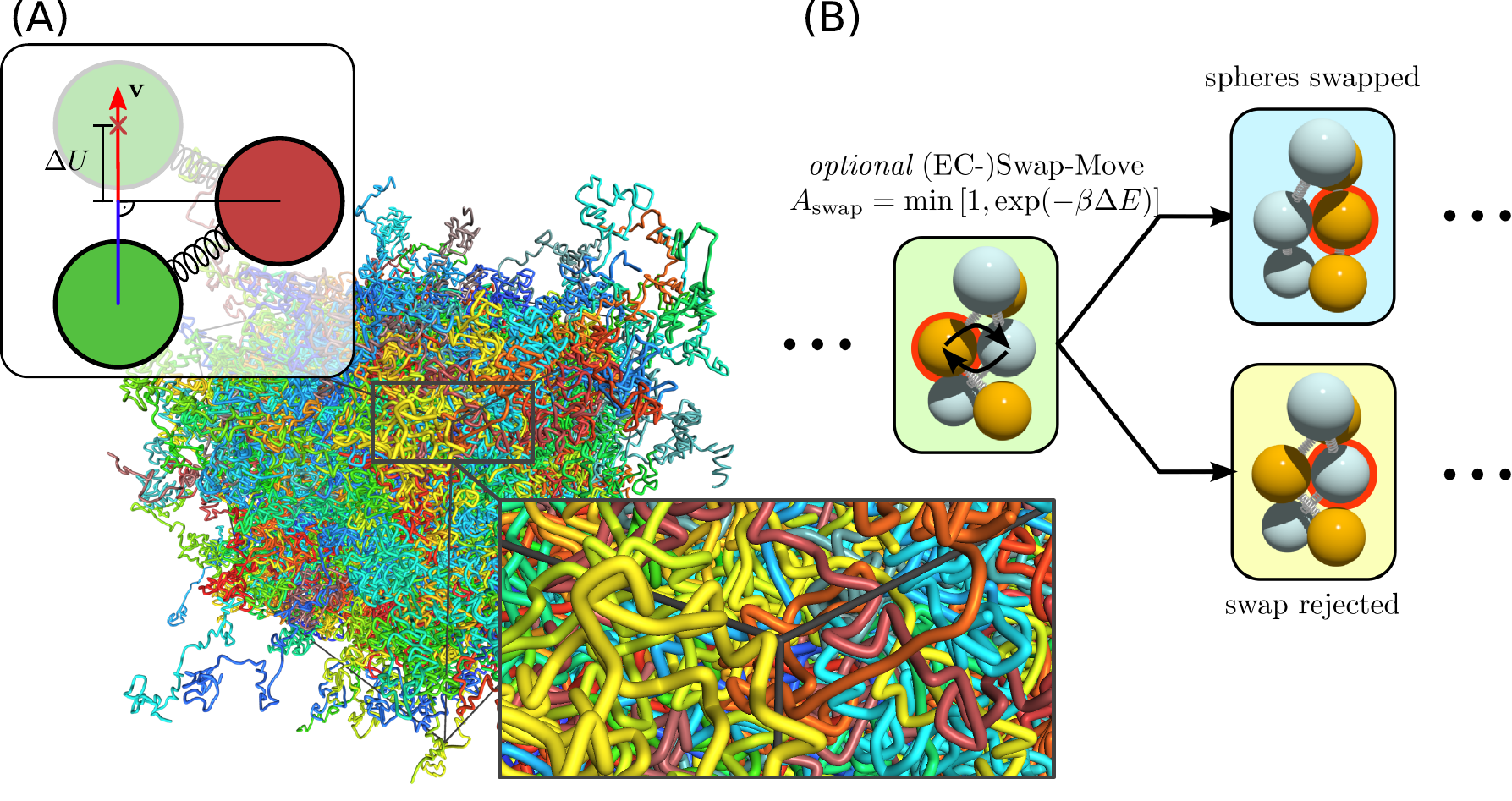}
 \caption{The background shows a polymer melt of flexible chains, which 
   are presented  as volumetric splines.  \textbf{(A)} Scheme for the rejection
   distance for hard sphere chains. If two bonded spheres do not collide, the
   active sphere can move freely until the minimal distance between both is
   reached. A \emph{usable} energy $\Delta E$ is drawn (see
   Fig.\ \ref{fig:2-EC}) which determines where the rejection takes
   place. \textbf{(B)} Scheme of a swap move. In case of a hard sphere
   collision, the energy difference of each involved springs is calculated
   when the beads would switch positions. The swap move is accepted with a
   standard Metropolis filter, otherwise the EC is continued
   normally. The active particle is marked by a red circle.  }
 \label{fig:melt_ref}
\end{figure}

The simulation of polymer melts by Molecular Dynamics (MD) or MC simulations
is a challenging problem, in particular, for long chains at high densities,
where polymers in the melt exhibit slow reptation and entanglement dynamics.
For chain molecules of length $N$ the disentanglement time increases
$\propto N^3\,$ \cite{doiedwards}, which makes equilibration of long chain
molecules in a melt a numerically challenging problem if only {\em local}
displacement moves of polymer segments are employed as in a typical
off-lattice MC simulation with fixed bond lengths
\cite{Curro1974,Baumgartner1981,Khalatur1984,Haslam1999} or fluctuating bond
lengths \cite{Rosche2000}, or in MD simulations
\cite{Kremer1990,putz2000,auhl2003}, 
 or for highly confined and therefore extremely packed melts\cite{ramos2018}.
In MD simulations, reptation dynamics
has been successfully identified in the entangled regime 
\cite{Kremer1990,putz2000,auhl2003,Hsu2016}.
In MC simulation, reptation dynamics has
been first identified in lattice models \cite{Kremer1983} or fluctuating bond
lattice models \cite{Paul1991}  with Rouse-like local bond moves.

In MC simulations,
{\em non-local} or {\em collective} MC moves can be introduced, for
example, chain-topology changing double-bridging moves
\cite{karayiannis2002}, which speed up
equilibration; dynamic properties, however, are no longer realistic
if such collective MC moves are employed.  Likewise,
reptation  can also
be introduced as explicit additional  local MC reptation move
\cite{Wall1975,Haslam1999} (slithering snake moves) to obtain faster
equilibration of a polymer melt but reptation dynamics is no longer
dynamically realistic then.

In order to apply the above scheme of  ECMC simulations to  polymer melts,
flexible polymers are modeled by a  bead-spring
model consisting of  hard spheres, which are 
connected by harmonic springs  \cite{kampmann2015}.
These are the only additional soft pair interactions in the ECMC
scheme for a flexible polymer melt.
We choose the bond  rest length to equal
the hard-sphere diameter $\sigma$ and a sufficiently large bond stiffness
such that bond-crossing becomes inhibited by the impenetrable
hard spheres and polymers remain entangled.

ECMC simulations can speed up MC simulations of polymers such that
simulation speeds become comparable to MD simulation speeds and
such that reptation dynamics can be clearly observed \cite{kampmann2015}.
In each EC move, all beads  that would collide successively
during a short time interval in a MD simulation are displaced at once.
This gives rise to a ECMC dynamics which is effectively very
similar to the realistic MD dynamics
and  statements about
   polymer dynamics are still possible.
  In Ref.\ \citenum{kampmann2015} we compared the ECMC algorithm to MD
 simulations
  implemented in  LAMMPS regarding the equilibration of polymer melts.
  We utilized the time, that the positional fluctuation of the
  most inner bead of each polymer needs to reach the diffusive regime, as gauge
  to introduce an ad-hoc interpretation of \emph{time} in the MC simulation
  (see Table I and Fig.\ 5 in Ref.\ \citenum{kampmann2015}).
  This showed that
    LAMMPS equilibrates more efficiently for moderate chain lengths. For
    long polymers ($N=500$) and when using
    an additional swap-move, ECMC simulations
    become equally
  performant, suggesting clear advantages for melts with even longer polymers.

The additional swap-move chanes the topology 
  of entanglements locally. 
In contrast to the  double-bridging move, which changes bonds, 
  the swap move changes topology by changing bead positions.
 For this purpose we  modify the EC move so
that the EC does not directly transfer to the next bead upon
hard sphere contact  but, instead, a {\em swap} of the two touching spheres 
is proposed, see  Fig.\ \ref{fig:melt_ref}.
Such an  additional swap move is EC specific and 
allows for a local change of entanglements.
An analogous swap move in a standard MC algorithm
needs to select pairs of beads such that the swap move
has a reasonable acceptance rate (the particles have to
be reasonably close). Moreover, the selection rule has to
satisfy detailed balance (for example, simply proposing
the nearest neighbor for swapping will lead to a violation
of detailed balance). Therefore, there is no straightforward
analogue of the EC swap move in a standard MC
simulation with local moves.

Here, we want to focus on another important aspect of EC moves
in polymer melt simulations. EC moves  can also be employed to
generate initial configurations that are already
representative of equilibrium configurations.
This keeps the equilibration or warm-up phase of any subsequent
simulation scheme (MC or MD) short. An early 
strategy to generate initial conditions is to slowly compress a very dilute
equilibrated melt, which can take a long time for very dense
target systems and becomes more difficult when chain lengths increase
\cite{Kremer1990}.
Another method that has been proposed first distributes  the chains without
interaction (phantom chains) and tries to resolve overlaps
in a way which distorts the chain statistics as least
as possible \cite{Kremer1990}. Over the years more
sophisticated ways of resolving overlaps have been proposed,
for instance a slowly increase of the 
strongly repulsive excluded volume potential \cite{auhl2003}.
This procedure is called
\emph{push-off} and the rate at which the potential is increased and the
potential range is extended determines to a large extent how much the original
statistics of the chains are disturbed.
The method can also be applied
to more realistic polymer models \cite{sliozberg2016}. 
Recent work by Moreira {\it et al.}  \cite{Moreira2015}
shows how the procedure of Auhl {\it et al.}
can be applied more efficiently and how the equilibration time can be
shortened  roughly by a factor of  $6$.
An alternative strategy that performs
even better  for very long chains (again by a factor of $\sim 5$)
has been introduced by Zhang {\it et al.}
\cite{Zhang2014}.
They employ a hierarchical approach, which 
uses sequential backmapping  from a coarse-grained representation in order
to re-introduce molecular details.

\begin{figure}
\centering
 \includegraphics[width=0.85\linewidth]{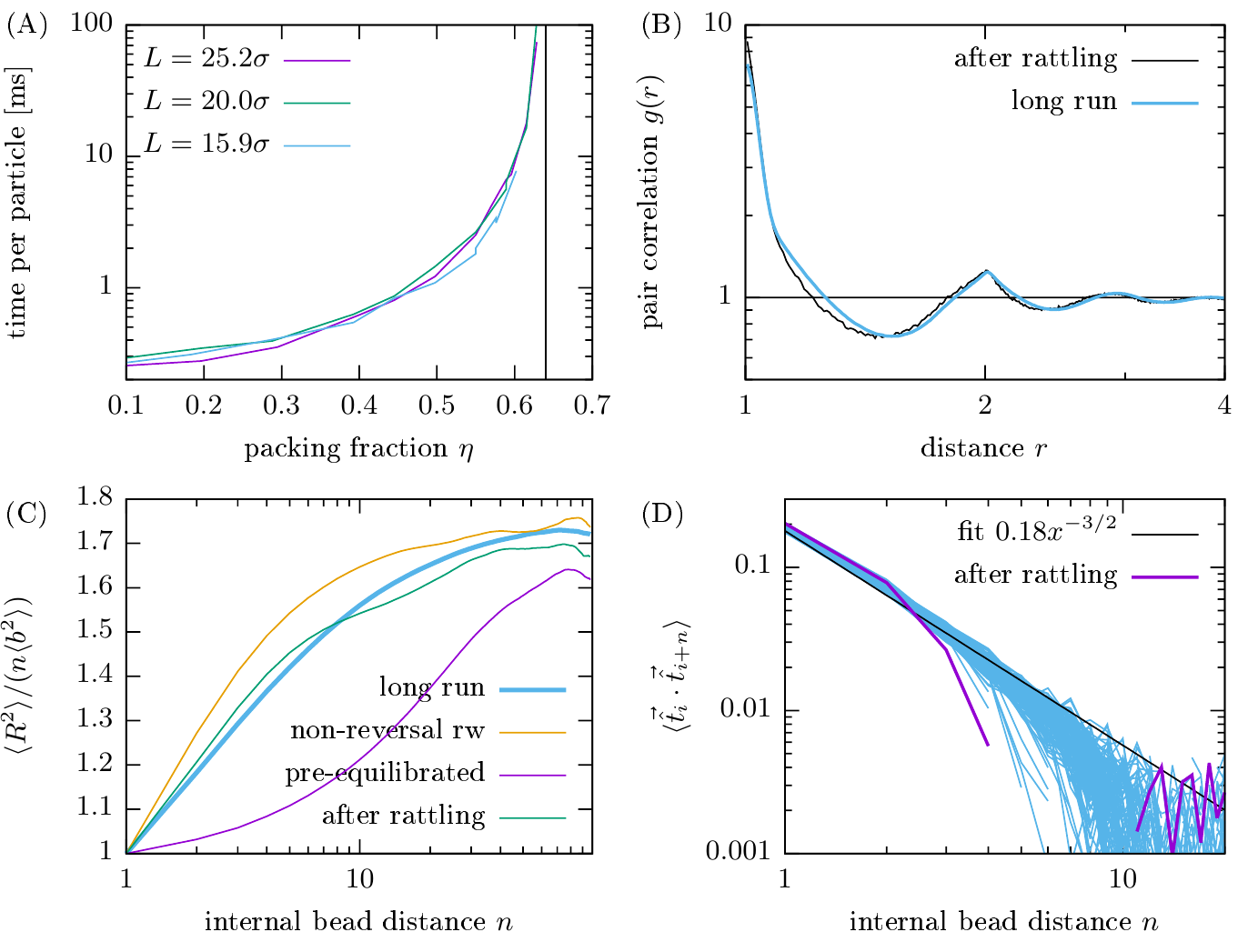}
 \caption{ \textbf{(A)} Wall time per particle to generate \emph{rattled}
   configurations of variable polymer number $M$
   with a fixed length of $N=200$.
   The wall time scales linearly in the numbers of particles,
   i.e., curves for different system sizes $L\times L$
   are identical and diverge when the
   packing fraction $\eta= MN (\pi/6) (\sigma^3 /L^3)$
   approaches the random closed packed limit
   $\eta_{rcp} \approx 0.64$ (indicated as vertical line). Preparation
    of even larger  systems
   ($M\times N=10^6$) for moderate $\eta$ ($\approx 0.4$) take only
   $\approx 10^3 \text{s}$. Preparation  of
     pure hard sphere systems ($N=1$) are roughly twice
   as fast. For the following we use $M=1000$ chains \`a $N=100$ beads at
   $\eta\approx 0.45$ ($\rho=0.85/\sigma^3$,
    i.e., $L\approx 49 \sigma$).
   All data is generated on a single core of
   a i7-8650U. \textbf{(B)} The pair correlation $g(r)$ is sensitive
   to the local structure
   of the hard spheres and resembles the equilibrium measurement quite
   well. The warm-up takes only a couple of minutes.
   \textbf{(C)} The internal
   bead-to-bead distance $\langle R^2\rangle / (n \langle b^2 \rangle)$
   captures how well the polymers are equilibrated and has to be as close to
   the equilibrium result
   as possible because this observable equilibrates very slowly.
   For equilibration 
    a chain needs to move its own size, which is slow due to reptational
   dynamics. Because of
   strong fluctuations the warm-up times have not been
   measured precisely, but are of the order of a couple of hours.
   \textbf{(D)}
   The tangent correlation shows the correct
   non-ideal behaviour of chains in the
   melt\cite{Wittmer2007,Wittmer2007a}.
    The configuration after \emph{rattling} resembles the equilibrium
   quite well. The warm-up takes only a couple of minutes.}
  \label{fig:melt_initial}
\end{figure}

We present an EC-based scheme that is similar to the push-off scheme and takes
advantage of the fact that the EC algorithm is well suited to resolve
overlaps. To do this, ECs are started in random directions on any sphere that
overlaps with at least one other sphere until all overlaps of this sphere are
resolved. This procedure ignores spheres that overlap with the active sphere,
so that the algorithm itself does not need to be changed.  This procedure
corresponds to locally \emph{rattling} the polymer system, and can create enough
space around the overlapping bead to insert it.

For relatively dense melts, Flory \cite{Flory1951,flory} puts forward the
hypothesis that the volume exclusion effect (chain swelling,
$\langle R^2 \rangle \sim N^{2\nu}$ with $\nu \geq {1}/{2}$) is exactly
compensated by the pressure generated by surrounding chains, so that a chain
shows ideal behaviour again. More recent results
\cite{Wittmer2007,Wittmer2007a} show that this hypothesis is only a good
approximation. In particular, the tangent-tangent correlation of long
($N > 500$) polymers in a melt falls algebraically, which deviates
significantly from exponential decorrelation happening in the single polymer,
i.e., free case.

This motivates the following scheme to generate pre-equilibrated polymer
melts. Since excluded volume interactions between parts of a given chain are
more or less screened as Flory suggests, we place phantom/ideal chains into
the system in configurations that resemble non-reversal configurations by
excluding a certain angular region for backward steps in generating the
initial phantom chain configurations as described in Ref.\ \citenum{auhl2003}.
This captures the structure of the chains on all length scales quite well, but
comes with the downside that the asymptotic behavior of the end-to-end distance
(the asymptotic Flory ratio $\langle R^2\rangle / (n \langle b^2 \rangle)$)
must be known beforehand to tune the excluded angular region accordingly.
Luckily, the internal structure of the chains
depends not strongly on the chain length and the asymptotic behavior can be
determined in a \emph{faster} system of shorter chains 
(here, we carried out one long run. Theoretical and numerical values for
  this ratio can be found in Ref.\  \citenum{foteinopoulou2008}
  and references therein).

The resulting phantom chain system is then equilibrated for a short time, so
that the phantom chains will relax towards ideal chain configurations, i.e.,
the phantom chains contract starting from the initial non-reversal
configurations.  Then the hard sphere diameter is increased from $\sigma=0$
(phantom chains) to the target $\sigma$ at once\footnote{In our previous work
  \cite{kampmann2015b} we suggested a steady increase of the hard sphere
  diameter (similar to the \emph{slow push-off}) rather than introducing the
  full diameter at once, but actually implemented
  the procedure outlined here because of an implementation error.
  Therefore, the procedure in Ref.\ \citenum{kampmann2015b} was
  described wrongly, but still produced overlap-free, high quality
  initial conditions.},
which gives rise to overlaps.  Then we use the EC
rattling moves to resolve overlaps in the system. Switching on the hard sphere
diameter will swell the chains and, therefore, have the opposite effect to the
pre-equilibration. Both steps cancel each other well, resulting in high
quality initial conditions. The quality can be assessed by comparison of
equilibrium values of observables, such as $\langle R^2\rangle$, pair
distribution $g(r)$, or bead distributions, with ensemble values given by the
initial condition generation as shown in Fig.\ \ref{fig:melt_initial}.

For this procedure it is vital that the spring constant is quite
large. Starting ECs on one particle until overlaps are resolved
obviously breaks
balance.  Presumably, setting the spring constant to a relatively high
value (in Fig.\ \ref{fig:melt_initial} $k= 1000 k_BT / \sigma^2$,
where the rest length is identical to the hard sphere diameter $b_0 = \sigma$)
leads to less distortions along the chain.

In general, it is hard to quantitatively
compare time scales and thus performance of
the EC-based initialization scheme with other schemes from the literature.
Hardware development over the years
(Moore's law, number of cores, etc.),
underlying polymer models (volume exclusion: hard spheres
vs.\ Lennard-Jones, bonding: harmonic springs vs.\ FENE vs.\
tangent hard spheres vs.\
united atom force fields), criteria of deciding when \emph{equilibrium} is
reached, or simply untested parameter dependencies
(regarding system size, chain length,
spring constants) can have non-trivial influence, even on an otherwise robust
benchmark. Despite these problems to compare quantitatively,
the proposed EC-based method seems
comparable to other state-of-the-art methods.
This has to be  checked further in future work.

\subsection{Self-Assembly Of  Filament Bundle Networks}

The cytoskeleton is a
mechanically important structure which consists of three classes of
interacting filamentous proteins (microtubules, filamentous F-actin,
intermediate filaments) in animal cells. The
different constituents reflect the multi-functionality of the cytoskeleton
and the different requirements with respect to force and time scales.
F-actin structures in the cell cortex
are most important for cell mechanics and cell motility
\cite{huber2013,blanchoin2014}. 
From the polymer physics point of view,
F-actin is  a semiflexible polymer 
since  typical contours lengths are  of the same order as its
persistence length. Therefore, bending rigidity and
thermal fluctuations are relevant to understand F-actin  structures. 
Within the cell,
F-actin  forms locally
very dense sub-cellular structures like bundles \cite{Bartles2000},
networks and even networks of bundles to
fulfill their biological functionality \cite{huber2013}.
\emph{In vivo}, bundles and also networks are held
together by crosslinking proteins \cite{Lieleg2010}.

Also 
minimal cytoskeletal \emph{in vitro} systems 
\cite{Lieleg2010}, some of which are 
confined to droplets \cite{huber2012,huber2015b}
or in microfluidic chambers \cite{deshpande2012,deshpande2015,strelnikova2016},
show formation of bundles and also networks of bundles under the
influence of attractive interactions. \emph{In vitro}, such
attractive interactions can not only be induced by crosslinkers
but also by counterions and depletion interactions.

The simulation of the cytoskeleton of
a cell is a long-standing problem as its physical properties are often
governed by the interplay of many long polymers, which are
deformable by thermal fluctuations against their bending rigidity and
subject to crosslinker-mediated attractive interactions \cite{Kierfeld2005}.  
Theoretical work on
crosslinker-mediated bundling of semiflexible polymers \cite{Kierfeld2003,
  Kierfeld2005, Kierfeld2005a, kampmann2013} and the related
problem of counterion-mediated binding
of semiflexible polyelectrolytes \cite{Borukhov2001} show a discontinuous
bundling transition above a threshold concentration of crosslinkers or
counterions.
The nature of the crosslinkers can influence the mechanical
properties of polymer bundles \cite{Shabbir2019}.

\begin{figure}
\centering
\includegraphics[width=\textwidth]{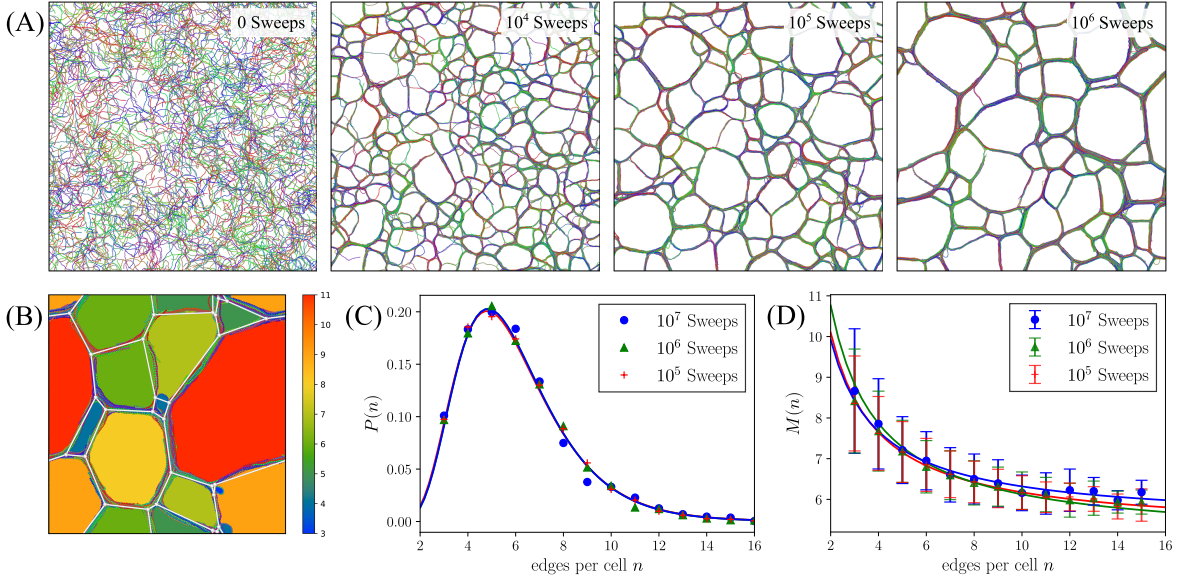}

\caption{ \textbf{(A)} A series of snapshots (after $0$, $10^4$, $10^5$, $10^6$
  sweeps; in a sweep, each sphere is moved once on average);
  demonstrating coarsening.
  The  snapshots are taken from a simulation of $M=1000$
  polymers \`a $N=100$ beads  in a 
    $400\,\sigma\times 400\,\sigma\times 10\,\sigma$ simulation 
    box. The simulation is performed with the following parameters:
     bending constant 
    $\kappa =30\, k_BT \sigma$, spring constant $k=100\, k_BT/\sigma^2$ and 
    rest length $b_0=\sigma$, potential strength 
    $g=0.7\, k_BT$ and range $d=0.4\,\sigma$.
    A random color is assigned to each polymer for clarity.
  \textbf{(B)} Example of a simple
  graph representation (white balls and bonds) as overlay over a network
  snapshot generated by an automated procedure to identify vertices (white
  balls) and edges (white bonds) of the bundle network.  The color of
  cells codes for  the number of
  enclosing edges. All data in Figs.\ C and D is measured from such graphs. For
  clarity, we use a comparatively small system.  \textbf{(C)} Distribution
  $P(n)$ of the number of edges $n$ per cell for different simulation
  times. The curves are obtained by fitting log-normal distributions to the
  data.  \textbf{(D)} Aboav-Weaire law $M(n)$ for different simulation times,
  which measures the mean edge counts of neighboring cells. Cells with a
  high number of edges tend to have neighboring cells with fewer edges and
  vice versa. Error bars represent one standard deviation.
  The Aboav-Weaire law, the fits of 
  $P(n)$ in C as well as direct computation of the
  variance $\mu_2$ of the number of 
  edges per cell yield similar values $\mu_2\sim 5.2$
   for the 
    simulation parameters given in A.
  }
 \label{fig:network}
\end{figure}

Simulations of few  filaments clearly confirm bundling
in a single transition \cite{Kierfeld2005}.
The emerging bundles of semiflexible
polymers (Fig.\ \ref{fig:network})
are typically rather densely packed which causes
difficulties in equilibrating bundled structures in traditional MC simulations
employing local moves. In Ref.\ \citenum{Kierfeld2005}, MC simulations showed
evidence for kinetically arrested states with segregated sub-bundles; in
Ref.\ \citenum{Stevens1999}, kinetically arrested bundle networks have been
observed for rather small semiflexible polyelectrolyte systems.
Further numerical progress requires a
faster equilibration of dense bundle structures.
Simulations of larger systems consisting of 
cytoskeletal filaments and  crosslinkers
\cite{Chelakkot2009, Cyron2013, foffano2016} show different
competing phases from isotropic networks to  bundles and other aggregates,
for example, with bond-orientational order.
In Ref.\ \citenum{foffano2016}, it was also pointed out that
the dynamics of bundling and polymerization in combination with
entanglement of  polymers strongly influence
the resulting structure. This raises the question to what extent
bundled structures can reach a genuine equilibrium under realistic
conditions. 

Further numerical progress requires a
faster equilibration of dense bundle structures, for which
the ECMC algorithm is ideally suited \cite{kampmann2015b}.
In comparison to polymer melts, two additional interactions
have to be included
in this system: (i) Polymers are semiflexible such that
we have to include a bending rigidity of each bead-spring polymer,
which is an example of a three-particle interaction  between
three neighboring hard sphere beads along a polymer.
(ii) There is a short-range attraction, which mimics a crosslinker-
or counterion-mediated attraction between polymers. 
Therefore, all beads 
also interact pairwise with a
short-ranged attractive square well
potential of strength $g$ and range $d$.
We choose a potential strength $g$ well above the critical value for bundling
\cite{Kierfeld2005,kampmann2013}, and 
the range of the attractive square well potential is $d = 0.4\sigma$,
i.e., comparable to the filament radius (bead radius).
Both additional interaction energies are incorporated in the ECMC
polymer simulation. Again, it is important to note that we expect
a qualitatively realistic MC dynamics from ECs,
which resembles collisions in MD simulations.
The performance can be further increases by parallelization 
as discussed above \cite{kampmann2015b}.

We have applied this ECMC algorithm to a system consisting of many\footnote{We
  tried as many as $M=5000$ polymers \`a $N=200$ beads \cite{kampmann2015}.
  Since the system should be large enough when the typical cell scale is much
  smaller than the system size, we use smaller systems to improve the
  statistics and to get to coarsening.} interacting
semiflexible polymers, such as actin filaments, in a flat simulation box in
three dimensions.  Periodic boundary conditions are used only for the
extended direction of the simulation box whereas the spatially short direction
is non-periodic. Under the influence of the short-range attraction, we
clearly observe formation of densely packed bundles.  We also observe that,
starting from an isotropic melt-like situation, it is not a single bundle that
forms but the system evolves into a structure consisting of a \emph{network}
of bundles.

The simulation does not reach a truly equilibrated stationary state, but the
network of bundles keeps evolving  by coarsening processes
as shown in Fig.\ \ref{fig:network}A.
This clearly demonstrates that the assembly dynamics
plays an important role in
structure formation in these systems, as similarly suggested in 
Ref.\ \citenum{foffano2016}.
As mentioned above, 
the  ECMC dynamics is effectively very
similar to the realistic MD dynamics. Therefore, the observed
coarsening dynamics is not an artefact of the MC simulation but
a generic property of this multi-filament system.
On the time scales of our simulations (up to $10^7$ MC sweeps)
  the bundle system keeps evolving by coarsening and no
  genuine equilibrium state is reached.

 Our preliminary results suggest that the bundle networks form
 structures similar to foams \cite{ohlenbusch1998}, which also
 continue to coarsen over time, see Fig.\ \ref{fig:network}.
 The dynamics of the coarsening
 process exhibits further analogies. It can be observed that
 cells in the bundle network, which enclose a comparatively large
 area, tend to increase their area at the expense of cells with
 smaller enclosed area. This behaviour
 is known from foams  as a
 consequence of the von Neumann law for the area growth rate
 of cells \cite{glazier1992} and of the
 correlation between relative bubble volume and the number of
 sides \cite{glazier1987}. However, the von Neumann law assumes
 an exchange of the enclosed medium by diffusion through the
 liquid interfaces and cannot be applied
 directly to the bundle networks considered here.
 Other empirical laws found in the context of foams like
 Aboav-Weaire law \cite{weaire1984} for the mean number $M(n)$
 of sides of
 cells surrounding an $n$-sided cell can also be applied very well to the
 emerging bundle networks (see Fig.\ \ref{fig:network}D). The 
 Aboav-Weaire law is of particular interest since it provides 
 an alternative approach to determine the variance $\mu_2$ of 
 the mean number of edges per cell (denoted as $P(n)$ in 
 Fig.\ \ref{fig:network}C). Comparison of the value of $\mu_2$ 
 computed directly from the data to the values obtained from 
 fit parameters of $P(n)$ and the Aboav-Weaire law yields a 
 reasonable agreement.

While the structure formation in dry foams is
driven by the minimization of interfacial  energies and diffusion
\cite{weaire2001}, 
it appears that the bundle  networks coarsen by a \emph{zipping}
mechanism \cite{Kierfeld2006a,kuhne2009} which leads – similar to the
decrease in interfacial energy in foams – to a decrease of the polymer binding
energy (i.e., to more adhesive contacts between polymers).
At a fixed amount of polymers this gives rise to a minimization of
bundle length, which plays an  analogous role to the minimization
 of surface area in a classical three-dimensional foam. 
The foam-like bundle networks resemble the structures observed
in different \emph{in vitro} experiments \cite{huber2012, deshpande2012},
where also  polygonal cell structures of the self-assembled
networks have  been observed.

\subsection{Liquid Crystals And Colloidal Suspensions}

\begin{figure}
\centering
 \includegraphics[width=0.95\linewidth]{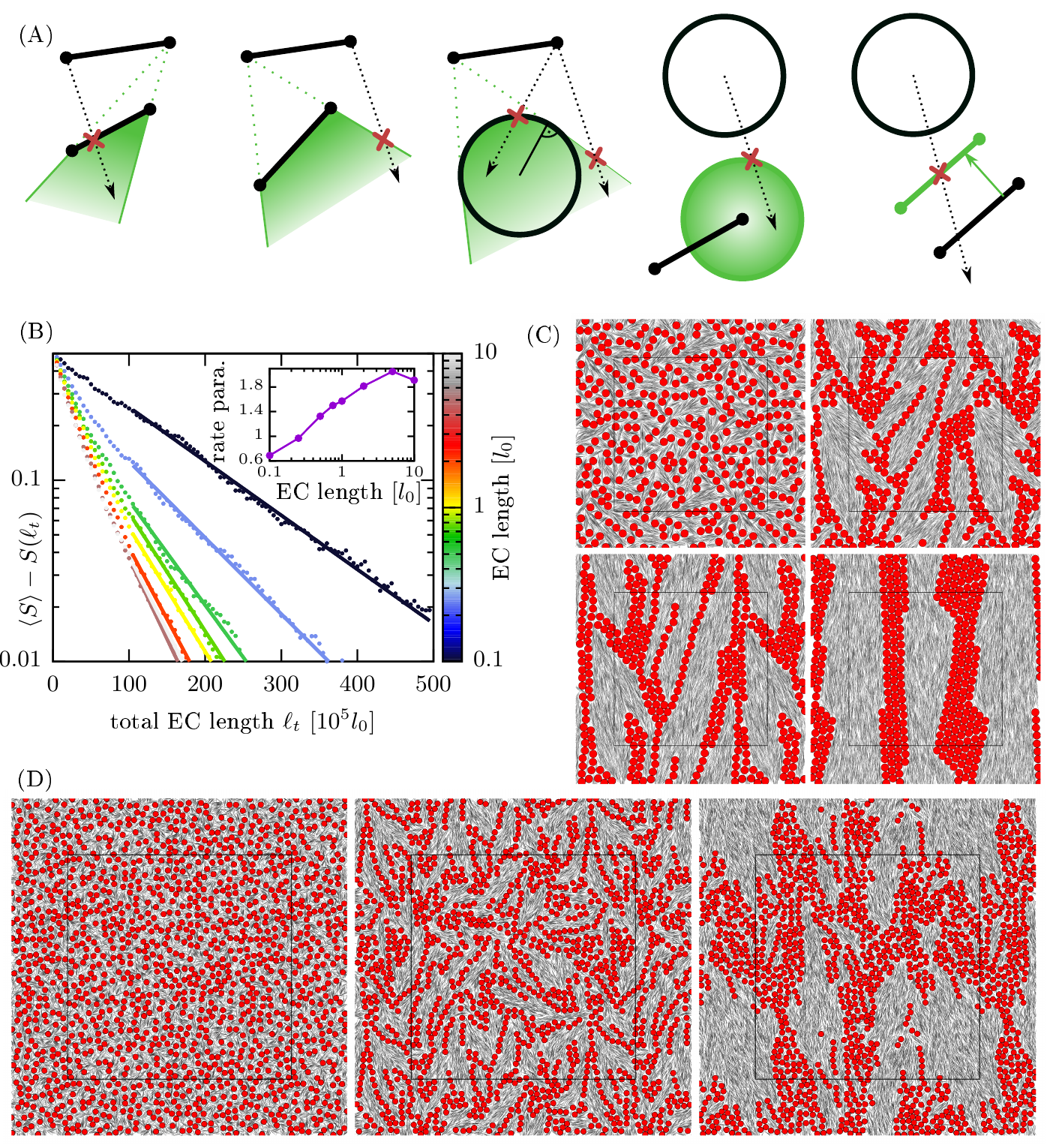}
 \caption{ \textbf{(A)} List of all geometrical cases where rejections/lifting
   between
   needles and disks occur (first two depict needle-needle, the third
   needle-disk, and the last two ones sphere-needle collisions). If a needle is
   not hit at one of its endpoints (first and last picture) conditional
   lifting probabilities $\lambda_{ij}$ for a three-particle interaction
    are applied.  
   \textbf{(B)} Efficiency
   gains with increasing EC length $\ell$ for a two-dimensional needle system
   in a volume of $L \times L$ at density $\rho_\text{n} = 10/l_0^2$.
   Starting at one random configuration an
   ensemble of simulations with different EC lengths are conducted. We show
   the difference between the nematic order parameter over time and the
   equilibrium value yielding an exponential. The decay rate of the
     order parameter is  used as
   efficiency measure, where a simulation with $\ell=5 l_0$ is $2.8$ times
   more efficient than a simulation with $\ell = 0.1$, where the latter took
   even $1.7$ times the wall time, yielding a speed-up of $\approx 5$ between
   $\ell=0.1$ and $\ell=5=L/2$. \textbf{(C,D)} Evolution of a mixture of
   hard needles of length $\l_0=1$ and
   hard disks with a diameter $\sigma=l_0$
    indicating an
    effective interaction between disks, which favors parallel chaining.
    \textbf{(C)} A system of size  $20\l_0 \times 20\l_0$
    containing  needles at density $\rho_\text{n} = 20/l_0^2$ and $160$
    disks of diameter $\sigma=l_0$,
    \textbf{(D)} a larger system of  $40\l_0 \times 40\l_0$
    containing  needles at density $\rho_\text{n} = 10/l_0^2$
    and  $640$ disks of diameter $\sigma=l_0$.}
 \label{fig:needle}
\end{figure}

Composite soft matter systems such as colloidal mixtures
containing different colloidal particles,
often of different size, represent challenging systems for simulations
because their physics are typically governed by effective interactions
which arise if microscopic degrees of freedom of
one species are integrated out. Effective interactions are
essential to characterize stability and
potential self-assembly  into crystalline phases but hard to access
in a microscopic particle-based simulation. The process of integrating
out microscopic degrees of freedom corresponds to the numerical evaluation
of a potential of mean force between the colloidal species of interest.
In order to measure the potential of mean force for one colloidal species
accurately, \emph{all}  degrees of freedom 
must be properly equilibrated.

As an example of such a colloidal mixture we consider
a two-dimensional system of needles and colloidal disks.
This serves as  a simple model system for liquid crystal (LC)  colloids,
which are colloidal particles suspended in a liquid crystal. 
 Such LC colloids exhibit
anisotropic effective interactions between colloidal particles
if the LC is in an ordered, e.g., nematic phase \cite{Stark2001,Lev2002}.
A nematic LC phase forms a rich variety of defect-structures
around a spherical inclusion  such as
\emph{Saturn-ring} disclination rings  or a satellite hedgehog
  for normal anchoring and boojums for planar
  anchoring \cite{Terentjev1995,Kuksenok1996,Ruhwandl1997a}.
  In a dense nematic LC, the effective interactions can be governed
  both by depletion interactions \cite{lekkerkerker2011} or by
  long-range elastic interactions mediated by director
field distortions in  the
nematic hard needles. Because hard needles
tend to align tangentially at a hard wall, we expect a quadrupolar
elastic interaction, which is characteristic for planar anchoring
at the colloidal disk
\cite{Ramaswamy1996,ruhwandl1997, mozaffari2011,tasinkevych2012,pueschel2016}
but also generic in two dimensions \cite{tasinkevych2002}. 
This interplay has been studied in Ref.\ \citenum{muller2020}
employing an ECMC simulation.

The colloidal mixture of hard disks suspended in a nematic host of hard
needles is particularly challenging as the hard needle system must be fairly
dense to establish a nematic phase. While particle-based simulations exist for
dilute rods in the isotropic phase\cite{schmidt2001,kim2004}, the regime of a
nematic host is fairly unexplored up to now and simulations resorted to
coarse-graining approaches.\cite{pueschel2016} So far, only single
inclusions\cite{Rahimi2017} or confining geometries\cite{Garlea2019} have been
investigated by particle-based simulations.
In order to measure the potential of mean force for larger colloidal disks 
accurately, all  degrees of freedom  of the needles but  also the relatively
slow degrees of freedom of the colloidal disks
must be properly equilibrated. This is efficiently
achieved by the ECMC algorithm.

The idea for the ECMC simulation of a needle system 
is to represent needles by their two endpoints, where only one endpoint is
moving at a time \cite{harland2017}; the
endpoints are connected by an infinitely thin, hard line.
For efficient sampling, the distance of the two endpoints, i.e. the
length $l$ of the hard needle can fluctuate around its
characteristic length $l_0$  in order to allow
for independent motion of both endpoints in the MC simulation;
the needle length $l$  is restricted by a  tethering  potential
$V_\text{n}(l)$
with  $V_\text{n}(l) = 0$ for  $l/l_0 \in [0.9, 1.1]$ and
infinite $V_\text{n}(l)$ else,  
such that $l_0$ is the effective needle length.
This constitutes an additional pair potential in the ECMC framework.

In two dimensions,
the needle-needle interaction simplifies
effectively to a collision of an endpoint with another needle. The remaining
MC move distance is lifted to one of the endpoints of this needle. Therefore,
we have a fluid of endpoints with an effective three-particle interaction (two
endpoints of a passive needle and the active endpoint).
For needles the probability $\lambda_{ij}$
to which endpoint is lifted (see eq.\ (\ref{eq:lambdaij})
is simply proportional to the
the distance to the other endpoint, i.e., it is lifted with higher probability
to the closer endpoint.

In a composite colloidal problem the ECMC algorithm also faces the problem
of lifting between different species. This can be performed exactly
according to the same rules as lifting between a single species. 
In the presence of additional disks, MC displacement of needle endpoints
is also lifted to disks if a needle collides with the disk and vice versa
as illustrated in Fig.\ \ref{fig:needle}A. 
The example of a needle system also shows that 
collision detection is often the computational bottleneck of ECMC
simulations. Here,  we use a
sophisticated neighbor list system to achieve high simulation speeds also in
the nematic phase.  Each particle is confined to a \emph{container}, which
triggers an event when the particle leaves it.  Then the neighbor list is
updated, which ensures that the neighbor lists are always valid. Particles are
added to the neighbor lists of the other particle and vice versa when their
containers overlap. This way, for different particles different container
shapes can be chosen. For the needles, a very narrow rectangle can be used,
which limits the computational effort for calculating the distances to the
next collision and makes the simulation significantly more efficient in the
nematic phase. In particular, the anisotropy of the needles can be assigned
particularly well without sacrificing any flexibility for the bookkeeping of
the spheres. Furthermore, we optimize the updating of lists by putting them
onto a collision grid.

In  Fig.\ \ref{fig:needle}B, we show a detailed benchmarking
for a two-dimensional pure needle system. 
With increasing EC length $\ell$, the
  sampling efficiency is increased, where an EC length comparable to
  the mean free path can be seen as the limit of a
  local  MC simulation. In
  Fig.\ \ref{fig:needle}B,  we relax an  isotropic initial configuration
  towards nematic equilibrium in a quite dense needle system at
  density $\rho_\text{n} = 10/l_0^2$. The resulting
  relaxation of the nematic order parameter is 
  exponentially, where we use the decay constant of the exponential
  as efficiency measure.  For EC lengths approaching the
  system size,  no further improvements occur as naively expected.

\begin{figure}
\centering
 \includegraphics[width=1\linewidth]{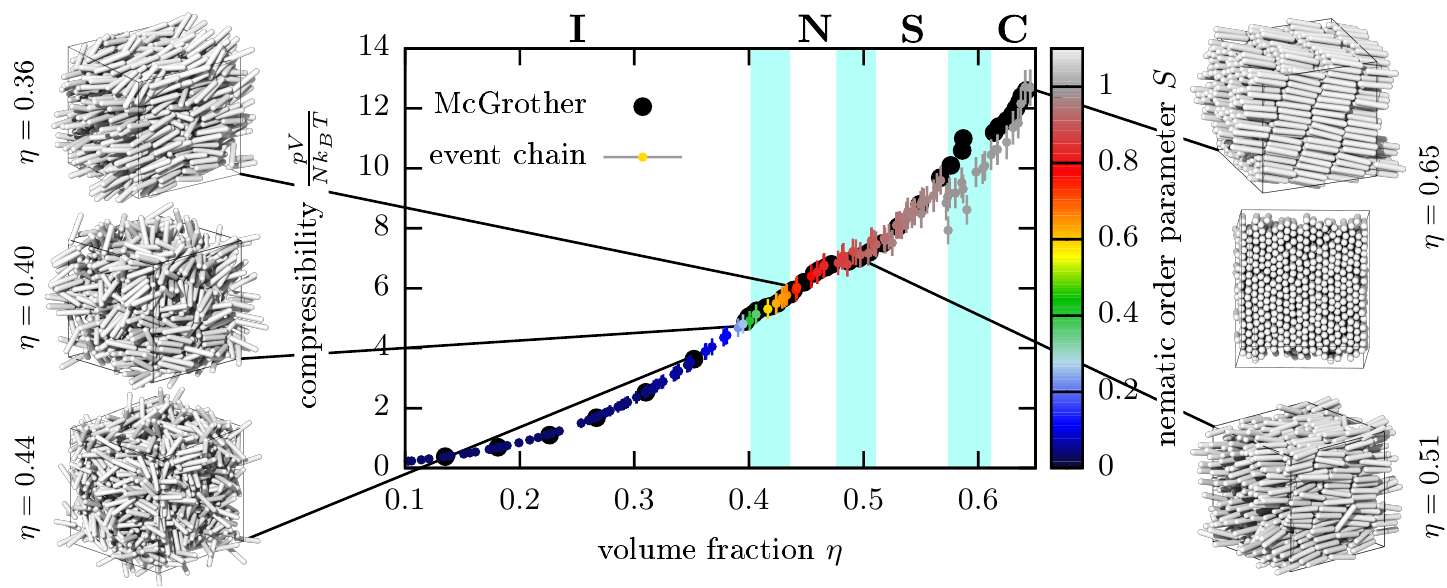}
 \caption{
   Phase diagram of hard spherocylinders (with diameter $D$
   and length $L=5 D$) in three dimensions. We see four phases
   ($\mathbf I$ - isotropic, $\mathbf N$ - nematic, $\mathbf S$ - smectic,
   $\mathbf C$ - crystalline), in agreement with
     previous results  by McGrother {\it et
   al.} \cite{mcgrother1996} We added six snapshots at different volume
 fractions $\eta$ to illustrate  the different phases and
  phase transitions (also indicated by the
  blue regions).
}
 \label{fig:sphero}
\end{figure}

  In three dimensions, hard needles introduce no excluded volume and
    therefore lack any phase transitions (we have checked with
    ECMC simulations not shown here that, independent of the density, the
    nematic order parameter vanishes for hard needles in equilibrium).
    In three dimensions 
    hard spherocylinders are a natural extension of needles that
    introduces a non-vanishing excluded volume.
    As an additional proof of concept of our
    ECMC implementation with three-particle lifting,
    we investigate the series of liquid crystal
    phase transitions by monitoring the pressure as a function of
    increasing volume fraction, see Fig.\ \ref{fig:sphero}.
    We see a sequence of transitions
    from  isotropic to  nematic ordering,  followed by a transition to
    smectic order, and ultimately, crystallization in complete  agreement
      with  literature results \cite{mcgrother1996}.

For the
  colloidal mixture of hard disks suspended in a nematic host of hard
needles,
the ECMC simulation reveals
a surprising and robust tendency for \emph{parallel}
chaining of disks along the
director axis which seems to contradict the chaining in a ${45^\circ}$ angle
with respect to the director axis as predicted by quadrupolar elastic
interactions in two dimensions.\cite{tasinkevych2012} This is a 
result of a dominant short-range depletion interaction,
which strongly favors parallel chaining \cite{muller2020}.
The effective disk-disk
interaction can  be obtained as a potential of mean force by
simulations of systems containing only two colloidal disks and is in
good agreement with analytical calculations, where we add the
depletion interaction mediated by needles on short scales
and the elastic quadrupolar interaction mediated by a nematic
needle LC which has weak planar anchoring at the colloidal disks and
exhibits elastic anisotropy.
Here, we demonstrate the validity of this results
on a larger scale for systems containing many colloidal disks.
Figures \ref{fig:needle}C and D show the  equilibration process of
160 and 640 disks with
diameter $\sigma = 1$ in needle densities of $\rho_\text{n} = 20/l_0^2$ and
$10/l_0^2$, respectively,
where $l_0$ is the equilibrium length of a needle. 
We clearly see that parallel chaining always corresponds to the
equilibrium state of the composite system.

\section{Discussion}

We presented several applications that demonstrate how
ECMC is an effective and fast simulation
technique that is particularly suited for dense soft matter systems.
Polymer melts, bundles of semiflexible polymers, and the composite
system of a liquid crystal colloid are computationally challenging
problems, where the ECMC techniques give an improved
performance. 

We started with an introduction to the algorithm, where we showed
how arbitrary interactions between hard spheres can be easily included into the
ECMC framework such that a multitude of soft matter systems can
be modeled. On this 
 algorithmic side, long-range interactions have not been
 covered explicitly here but can also be included in an effective
 manner \cite{Kapfer2016}.

Future developments
will also address variations of the basic EC moves presented here in order to
sample even more efficiently.
On a generic collision, one interaction rejects the move and lifting occurs
which means the current EC direction has a component parallel to the
energy gradient of this specific interaction. One can resample the
perpendicular parts of the EC direction in order to avoid reshuffling the EC
direction from time to time, i.e., a finite EC length. This procedure has
been  recently developed and is called forward EC \cite{michel2020}.
In specialized settings, the
scheme seems very promising with regard to efficient sampling
and should also be tested in the context of soft matter systems in the
future. 

On the application side in the soft matter field,
we presented new results for the initialization and pre-equilibration of
polymer melts, where a novel \emph{rattling} algorithm based on
ECs can be employed to efficiently remove overlaps in initial configurations.
For large systems of semiflexible polymers with a short-range attraction,
we demonstrated that ECMC simulations are capable to follow the formation
of large networks of bundles and their subsequent foam-like coarsening
behavior, see Fig.\ \ref{fig:network}.
For a two-dimensional hard disk and needle model for LC colloids,
the structure formation by parallel chaining of disks could be
followed for large systems over large time scales as shown
in Fig.\ \ref{fig:needle}. Here we find that increasing the EC length, i.e.,
increasing the \emph{cluster size} in our EC algorithm
gives rise to considerable performance gains.

In summary, we addressed compact quasi-zero-dimensional hard sphere particles,
one-dimensional hard rods or hard needles, and one-dimensional
polymers as constituents in soft matter ECMC simulations.
A natural extension for future work is the development of
ECMC algorithms for 
two-dimensional triangulated surfaces  with local elastic properties,
which are impenetrable for other
simulation objects, e.g., hard spheres or  polymers.
 This will allow us to study
fluctuating triangulated elastic membranes and, in a compound polymer-membrane
system, polymer networks confined to elastic capsules. It was shown that the
boundary conditions significantly contribute to the network
structure \cite{deshpande2012}. The elastic properties of a triangulated
surface can be described by a set of elastic energies such as the TRBS
model \cite{delingette2008}, where elastic constants like Young's modulus
and Poisson ratio are adjustable and also
bending stiffness can be included \cite{tamstorf2013}.

The actual implementation will be a very challenging problem.
 The interaction of a triangle with a point or a hard sphere is an effective
4-particle interaction. The rejection distance is efficiently calculable
by the M{\"o}ller–Trumbore intersection algorithm.
To implement non-intersecting surfaces, 
interactions between two triangles have to be introduced, which
can lead to  a  6-particle interaction.
Therefore, membrane simulations would also represent a  
further test of the general framework for  $\mathcal{N}$-particle
interactions in the 
EC algorithm \cite{harland2017}. The conditional
lifting probabilities will look similar to the case of hard needles.

Based on the extensive experiences with EC based simulations presented here,
we think
this simulation technique is suitable to tackle large (biological) systems. In
order to verify and evaluate the EC algorithm for triangulated
surfaces, we can investigate the crumpling transition \cite{kantor1987}. In
Ref.\ \citenum{weichsel2016}, the necessary biophysical conditions for the
formation of tubular membrane protrusions have been investigated and, it has
been suggested that actin filaments polymerizing against a soft membrane are
sufficient to form protrusions, which seems a suitable next system to explore
the possibilties of EC algorithms for dense soft matter systems.

\end{document}